\documentclass{emulateapj}
\usepackage{apjfonts}

\slugcomment{}

\shorttitle{RCW86 with {\it Suzaku}}
\shortauthors{Tsubone et al.}

\begin{document}

\title{%
A Systematic Study of the Thermal and Nonthermal Emission
in the Supernova Remnant RCW~86 with {\it Suzaku}
}%
\author{Yoshio Tsubone\altaffilmark{1},
Makoto Sawada\altaffilmark{1},
Aya Bamba\altaffilmark{1,2,3},
Satoru Katsuda\altaffilmark{4},
Jacco Vink\altaffilmark{5}
}
\altaffiltext{1}{
Department of Physics and Mathematics, Aoyama Gakuin University
5-10-1 Fuchinobe, Chuo-ku, Sagamihara, Kanagawa 252-5258, Japan
}
\altaffiltext{2}{
Department of Physics, The University of Tokyo,
7-3-1 Hongo, Bunkyo-ku, Tokyo 113-0033, Japan
}
\altaffiltext{3}{
Research Center for the Early Universe, School of Science, The
University of Tokyo, 7-3-1 Hongo, Bunkyo-ku, Tokyo 113-0033, Japan
}
\altaffiltext{4}{Department of Physics, Faculty of Science \& Engineering,
Chuo University,
1-13-27 Kasuga, Bunkyo-ku, Tokyo 112-8551, Japan
}
\altaffiltext{5}{Anton Pannekoek Institute/GRAPPA, University of Amsterdam
PO Box 94249, NL-1090 GE Amsterdam, the Netherlands
}

\begin{abstract}
Diffusive shock acceleration by the shockwaves in supernova remnants (SNRs) is widely accepted as 
the dominant source for Galactic cosmic rays. However, it is unknown what determines 
the maximum energy of accelerated particles. The surrounding environment could be one of the key parameters.
The SNR RCW~86 shows both thermal and non-thermal X-ray emission with different spatial 
morphologies. These emission originate from the shock-heated plasma and accelerated electrons respectively,
and their intensities reflect their density distributions. Thus, the remnant provides a suitable 
laboratory to test possible association between the acceleration efficiency
and the environment.
In this paper, we present results of spatially resolved spectroscopy 
of the entire remnant with Suzaku. The spacially-resolved spectra are well reproduced with a combination of
a power-law for synchrotron emission and a two-component optically thin thermal plasma, corresponding to the shocked
interstellar medium (ISM) with {\it kT} of 0.3--0.6~keV and Fe-dominated ejecta.
It is discovered that the photon index of the nonthermal component becomes smaller
with decreasing the emission measure of the shocked ISM,
where the shock speed has remained high.
This result implies
that the maximum energy of accelerated electrons in RCW~86 is higher in the low-density
and higher shock speed regions.
\end{abstract}


\section{Introduction}
Cosmic rays are particles which bombard the Earth from anywhere (\citealt{Hess}). X-ray and
GeV/TeV $\gamma$-ray observations have revealed that the shocks of supernova remnants (SNRs)
are the acceleration sites of the Galactic cosmic rays up to the TeV range
(e.g., \citealt{1995Natur.378..255K}; \citealt{2013Sci...339..807A}).

The diffusive shock acceleration mechanism (DSA: e.g. \citealt{2001RPPh...64..429M})
is believed to be the
relevant mechanism, which can explain the power-law energy distribution. However it is still
unclear what determines the efficiency. The surrounding environment should be one of the key
parameters (\citealt{2009ApJ...696.1956P}, for example). Typical X-ray spectra of SNRs show
either synchrotron emission from high energy electrons, thermal emission from shock-heated
interstellar medium (ISM), or combination of these. Therefore, by comparing spatial distributions
of these emission, we may reveal association between the acceleration efficiency and the
environment.

RCW~86 is a SNR located at (l, b) = (315.4, $-$2.5) with an apparent diameter of $\sim$42~arcmin.
The distance is estimated to be $\sim$2.8~kpc (\citealt{1996A&A...315..243R}).
\citet{1977ApJ...213L..53N} discovered the X-rays from RCW~86 for the first time. Because of
the low plasma temperature ($\sim$0.5~keV), RCW~86 had been assumed to be an old remnant. However,
since the detection of high temperature emission by Chandra and XMM-Newton, RCW~86 is accepted as
the remnant of SN~185 (\citealt{2006ApJ...648L..33V}). The average radio spectral index is 0.6
(\citealt{1988Ap&SS.148....3G}). Furthermore, the X-ray spectra of the SNR show not only thermal
X-rays but also synchrotron X-rays (\citealt{2000PASJ...52.1157B, 2001ApJ...550..334B}),
implying that this SNR is an electron accelerator up to $\sim$TeV range,
together with gamma-ray detection from GeV to TeV band
\citep{2014A&A...567A..23Y,2009ApJ...692.1500A}.
In fact, evidence for effective cosmic-ray acceleration has been reported
(\citealt{2005ApJ...621..793B, 2011ApJ...737...85H, 2013MNRAS.435..910H}),
although \citet{shimoda2015} suggested the measured efficiency could be higher than reality.
It is noteworthy that the thermal and non-thermal emission in this SNR is not uniform and have
different morphology, which makes this SNR a suitable laboratory to test a possible relation
between the acceleration efficiency and the environment.

In this paper, we present the systematic analysis of spatially resolved spectra obtained with
the X-ray Imaging Spectrometer (XIS: \citealt{2007PASJ...59S..23K}) onboard Suzaku
(\citealt{2007PASJ...59S...1M}). Unless otherwise noted the uncertainty is given at the 90\%
confidence interval.

\section{Observation and Data Reduction}

\begin{figure}
\epsscale{0.8}
\plotone{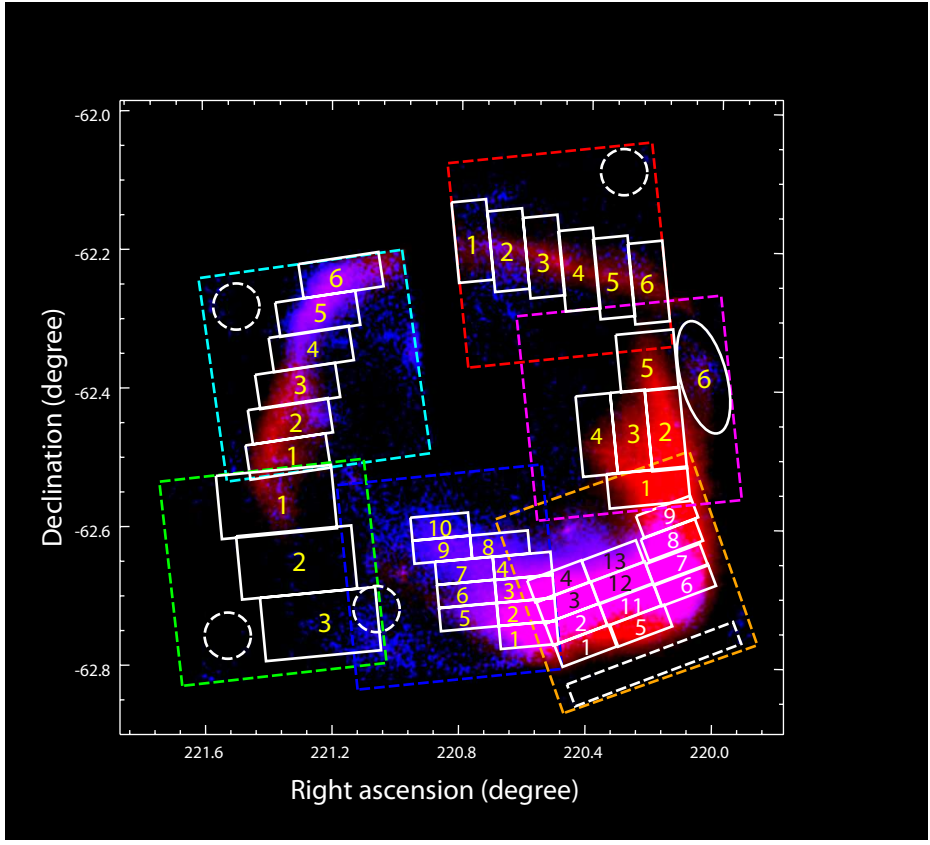}
	\caption{The XIS image of RCW~86 in a logarithmic scale. Red and blue respectively show
	emission from 0.5--2.0~keV (soft) and 3.0--5.5~keV (hard). The FoVs of XIS are indicated
	by the dashed squares. They are EAST (cyan), SE (green), SOUTH (blue), SW (orange),
	WEST (magenta) and NORTH (red). The solid-white shapes are source regions, while the 
	dashed-white shapes
	are background regions in each FoV.
The white and black numbers in the SW region represent SW-A and SW-B regions, as shown in the main text.
}
	\label{figure:image}
\end{figure}

The entire region of RCW~86 is covered by six observations with Suzaku. The observation log 
is summarized in table \ref{table:data}. The fields of views (FoVs) are shown in 
Fig. \ref{figure:image} with the dashed squares.

The XIS consists of four X-ray CCD cameras, which observes in 0.2--12.0~keV with X-ray
telescopes (XRT: \citealt{2007PASJ...59S...9S}). Three of them (XIS~0,~2, and~3) are
front-illuminated (FI) CCDs, and the other (XIS~1) is a back-illuminated (BI) CCD. The FI
CCDs have larger effective area in the hard X-ray band while the BI CCD in the soft band. A pre-flight energy
resolution is $\sim$ 130~eV (Full width at half maximum, FWHM) at 5.9~keV
(\citealt{2007PASJ...59S..23K}). After 2006 November 9, the XIS~2 was out of operation. The
others (XIS~0,~1, and~3) use spaced-row charge injection (SCI: \citealt{nakajima2008,2009PASJ...61S...9U}) to
recover the charge transfer efficiency and improving the energy resolution after 2006 October.
Each XRT accumulates X-rays with the effective area of 250~cm$^2$ at 8.0~keV to make an
image over an $18^{\prime}\times18^{\prime}$ FoV with angular resolutions of
$1^{\prime}.9$--$2^{\prime}.3$ in the half-power diameter.
Thanks to the low-Earth orbit, the non-X-ray background (NXB) is fairly low and stable.
Thus, Suzaku is very suitable for RCW~86, with its large extent and low surface brightness in some regions.

We reprocessed the data using the xispi software and the recent calibration database
released on 2012 May 9. The data were reduced in accordance with the standard screening
criteria\footnote{see (http:\slash{}\slash{}heasarc.nasa.gov\slash{}docs\slash{}suzaku\slash{}processing\slash{}criteria\_xis.html)}.
Grades 0, 2, 3, 4, and 6 events were used in following analysis. We removed events during the
South Atlantic Anomaly passages, Earth elevation angles below 5$^{\circ}$, and Earth day-time
elevation angles below 20$^{\circ}$. We also removed hot and flickering pixels. The
resultant systematic uncertainty in the energy gain is $<$ 12~eV. The XIS was operated in the
normal clocking mode and full-window mode in the all observations.

In this paper, we use the software packages HEAsoft version 6.10 and XSPEC version 12.7.1
and AtomDB version 1.3.2.
In the spectral analysis, we generated the redistribution matrix function and ancillary response
file using xisrmfgen and xissimarfgen (\citealt{2007PASJ...59S.113I}), respectively.

\section{Results}

\subsection{Image}
In Figure~\ref{figure:image}, we show the XIS images of RCW~86. Red and blue show the 
0.5--2.0~keV and 3.0--5.5~keV band images, respectively. The soft band is dominated by the thermal
emission, whereas the hard one is dominated by non-thermal emission. NXB events were generated by xisnxbgen
(\citealt{2008PASJ...60S..11T}) and subtracted from the image. We corrected exposure and
vignetting using the flat images at 1~keV (soft band) and 4~keV (hard band) generated by xissim
(\citealt{2007PASJ...59S.113I}). Both of these images show similar
shell-like structure. However, their distributions are not identical. They show offsets in the
azimuthal direction. For example, in the southern half of the shell, the soft emission appears
from west to southwest, while the hard emission from southwest to south. They also show difference in the
radial distributions. The soft emission is sharper while the hard emission is broader and
patchier. These results are consistent with previous observations (e.g., ASCA: 
\citealt{2000PASJ...52.1157B, 2001ApJ...550..334B}, Chandra: 
\citealt{2005ApJ...621..793B, 2002ApJ...581.1116R, 2013ApJ...779...49C},
XMM-Newton: \citealt{2006ApJ...648L..33V, 2014MNRAS.441.3040B}).

\subsection{Spectral Analysis}
As the image shows, the thermal and non-thermal emission have different spatial
distribution. For the spatially-resolved spectroscopy we divided RCW~86 in regions and determined physical
parameters using the following steps. First, we determined the parameters such as temperature for
each FoV (see Section \ref{eachFoV}). Next, we subdivided the FoV in regions, and 
we fitted the spectra of these regions using the
best-fit model for each FoV, but the normalization and power-law index were treated as free
parameters (see Section \ref{regions}).
Generating the ARF file, we assume the 1.7--5.0~keV image as the brightness
distribution. We used the spectral data of XIS-FI CCDs.

\subsubsection{Integrated spectra in each FoV}\label{eachFoV}

\begin{figure}
\epsscale{0.4}
\plotone{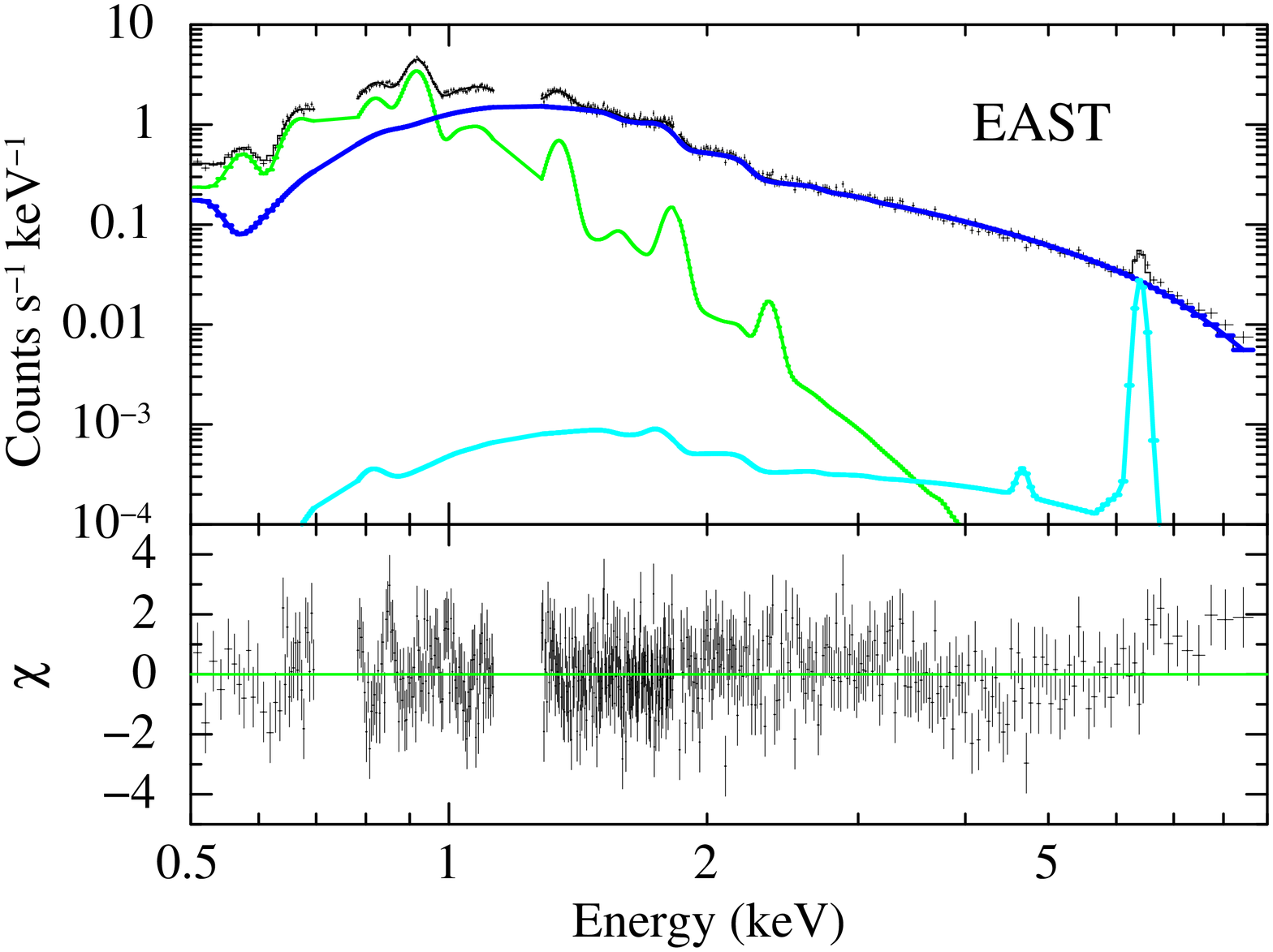}
\plotone{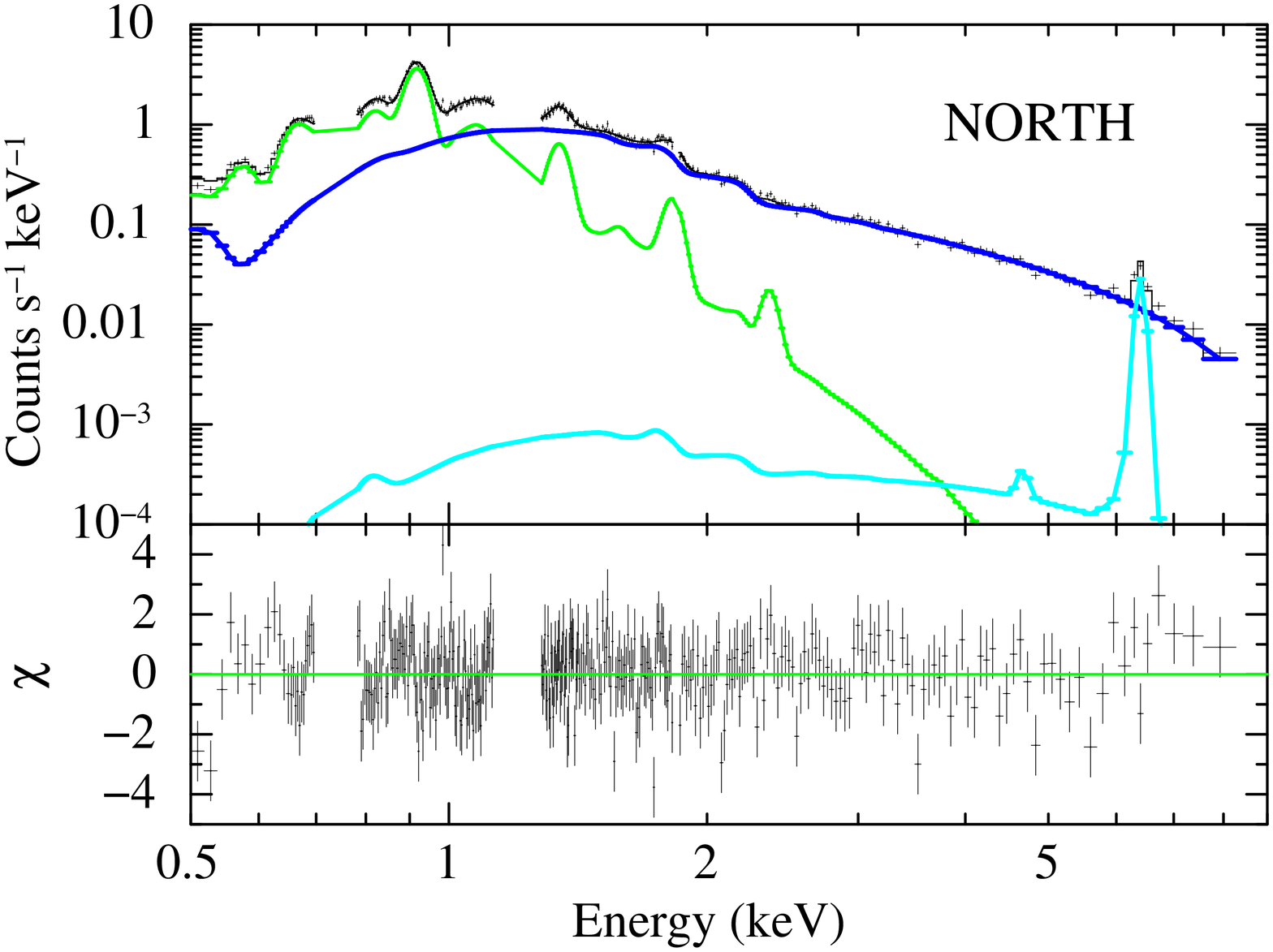}
\plotone{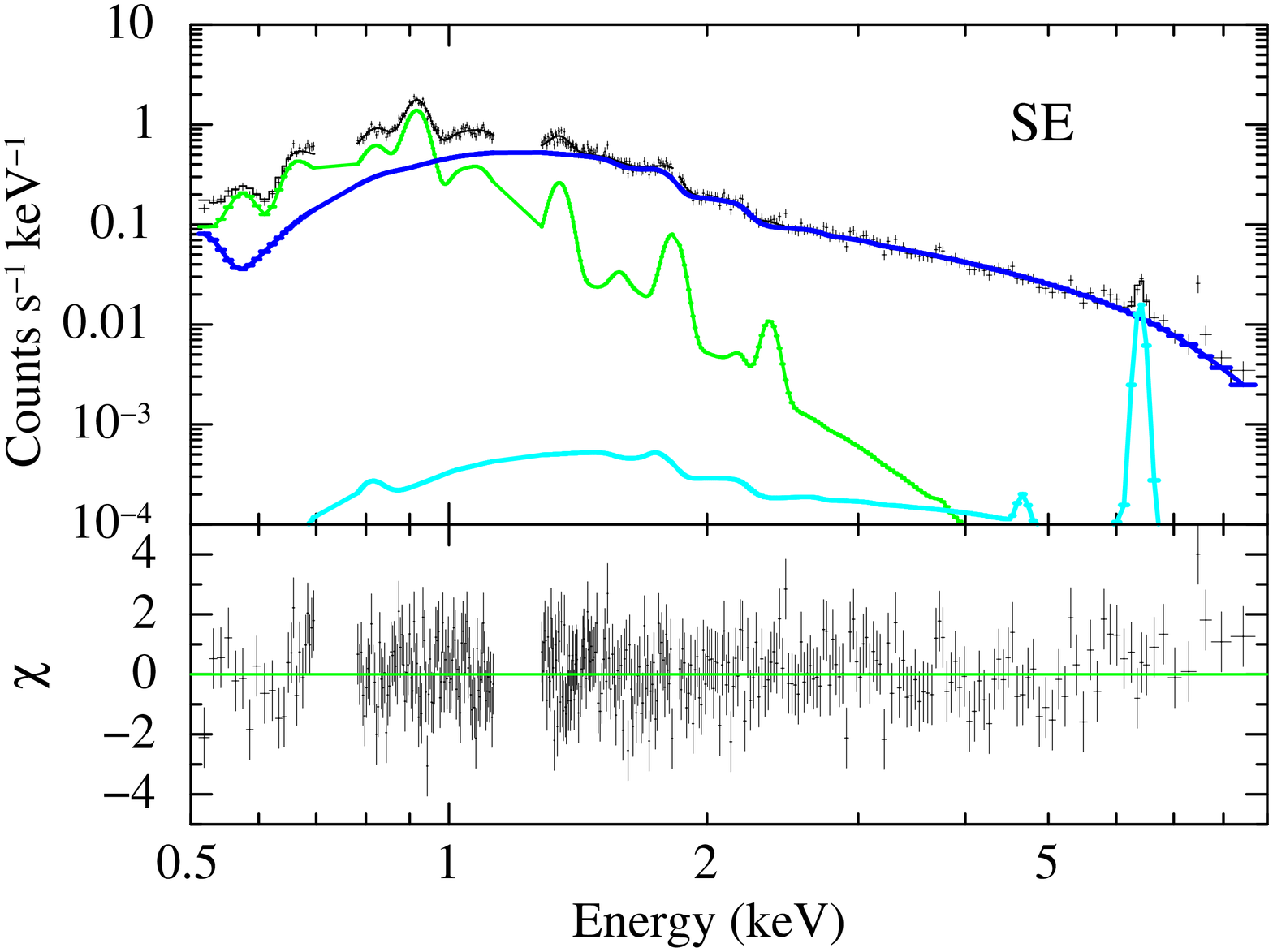}
\plotone{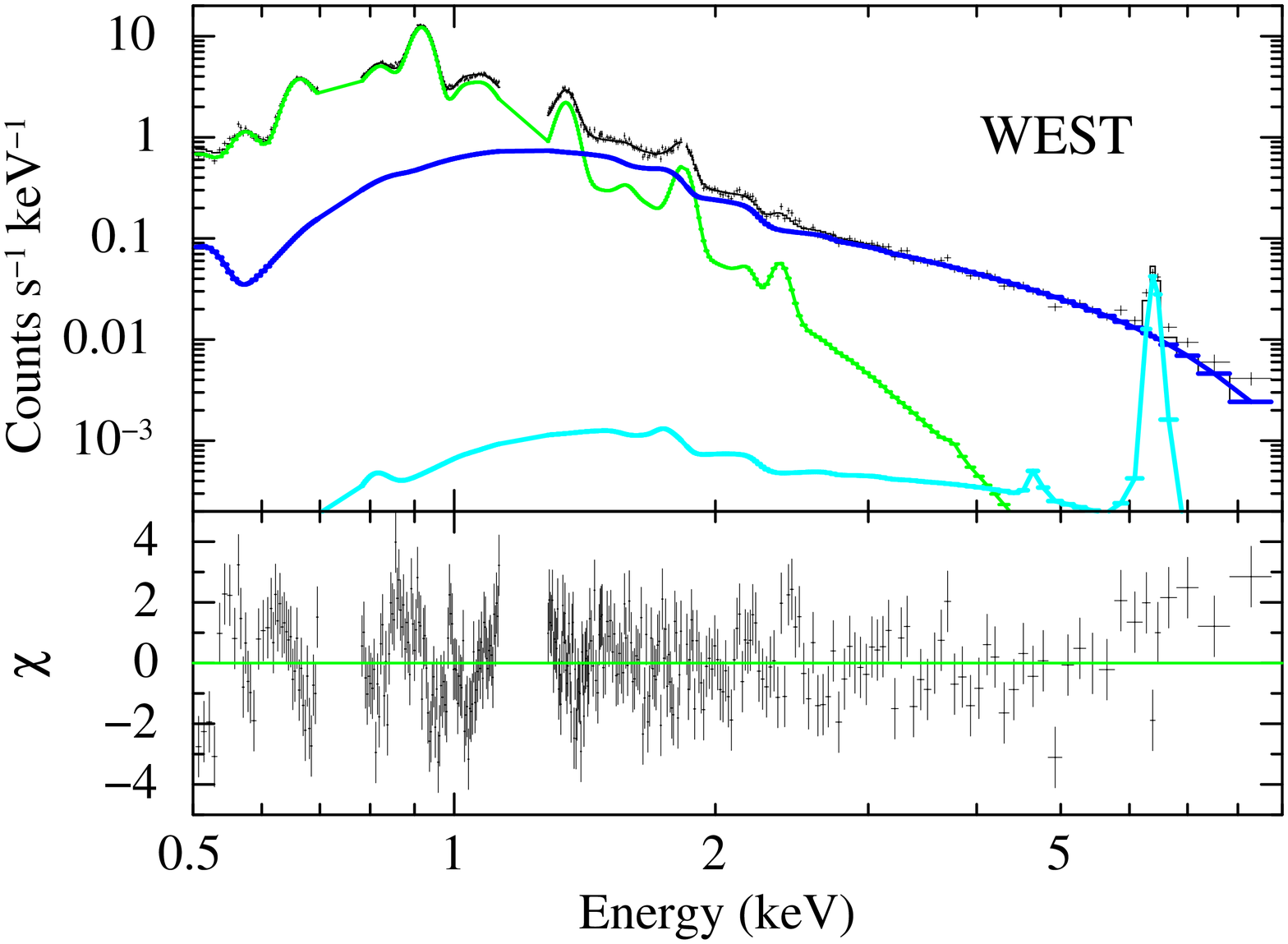}
\plotone{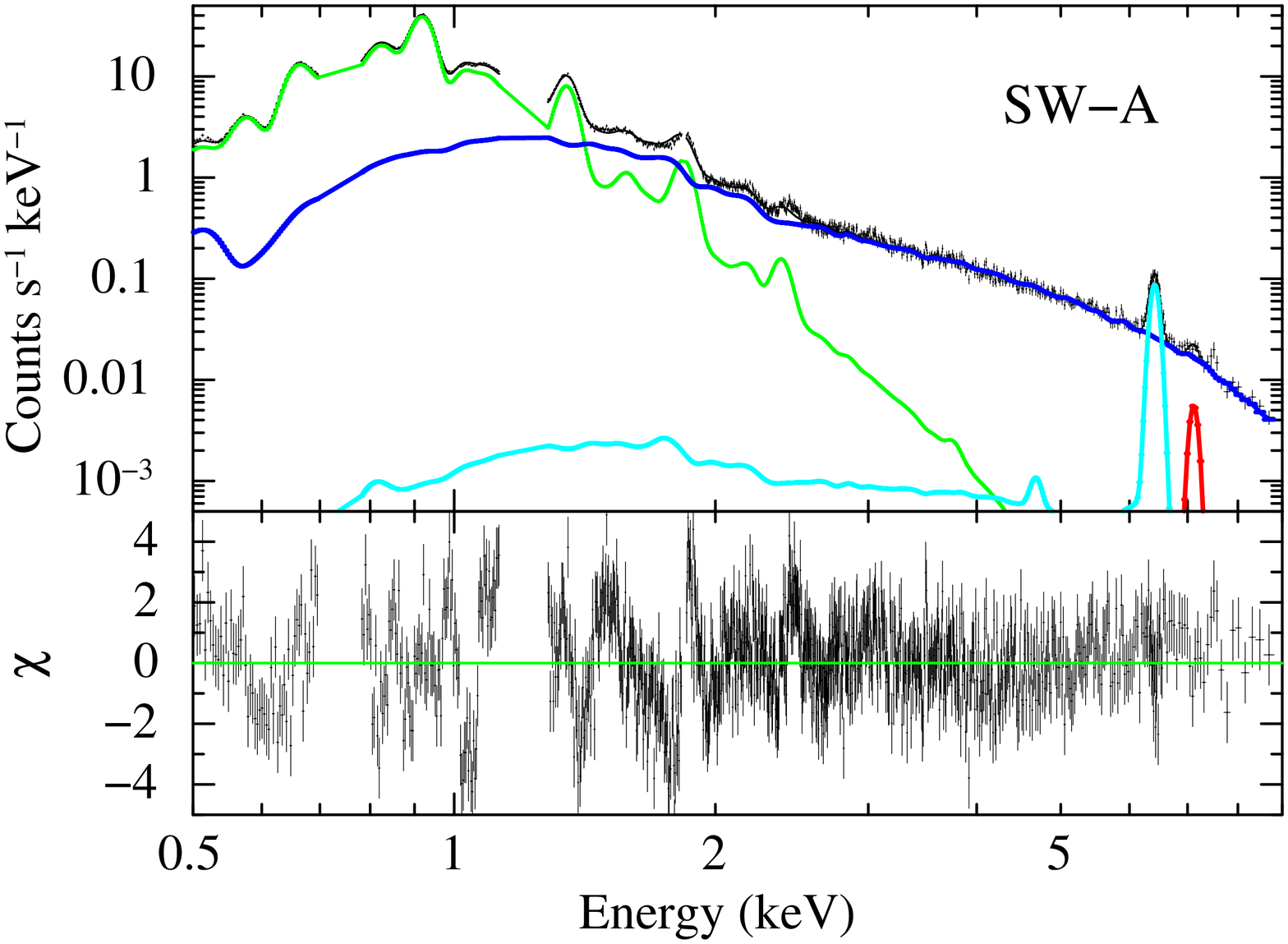}
\plotone{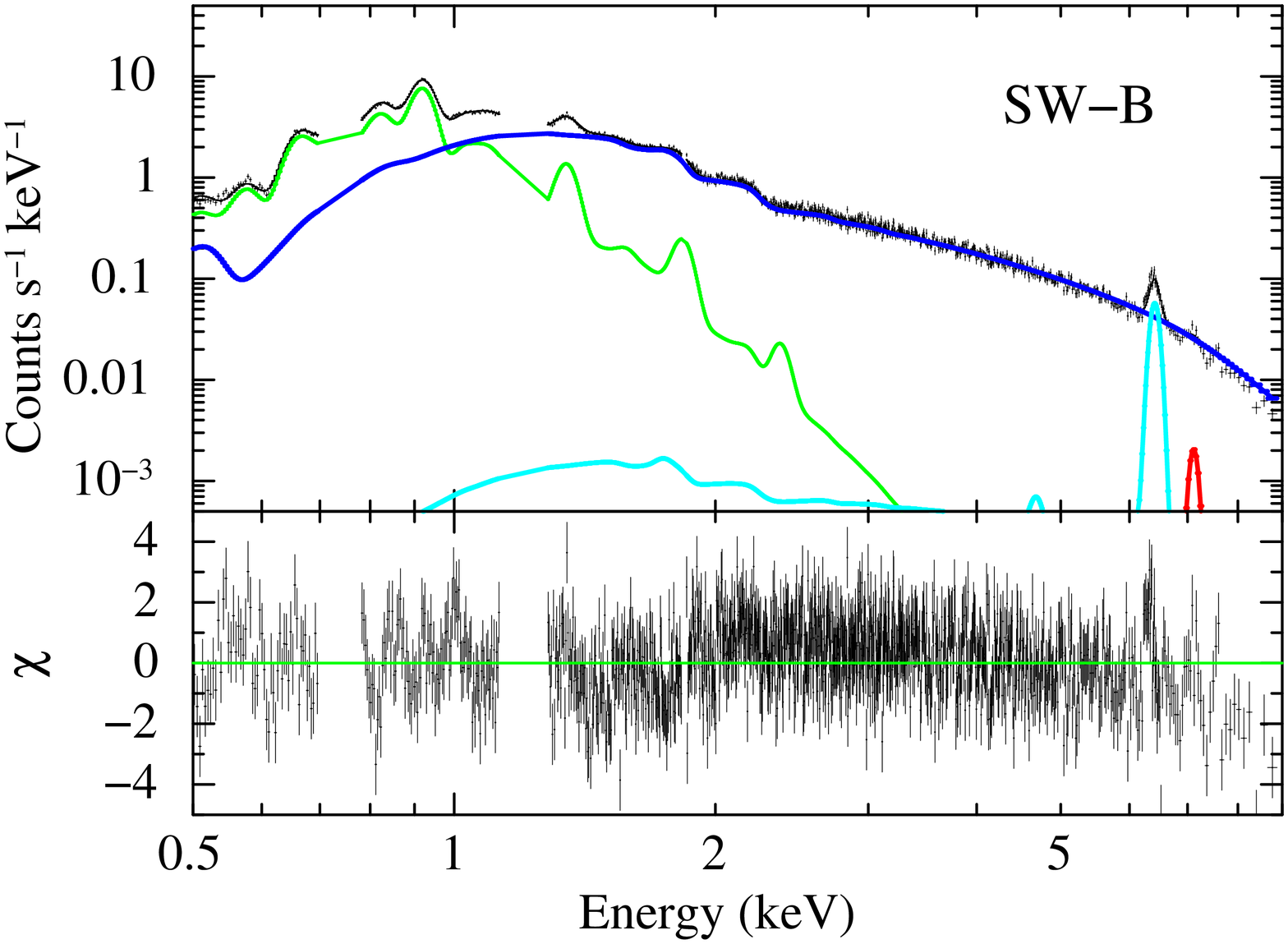}
\plotone{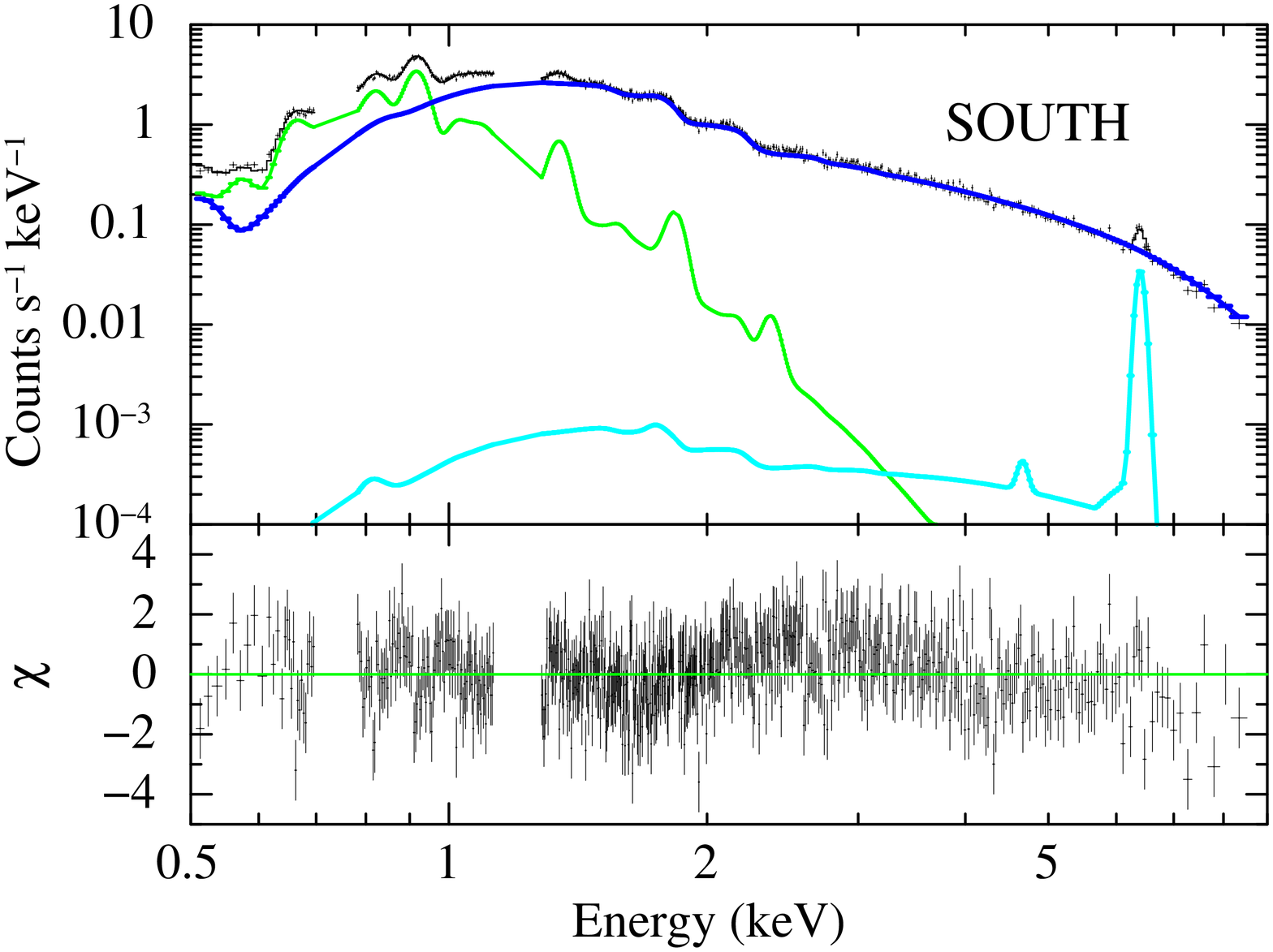}
	\caption{The Integrated spectra in each FoV. The blue, green, and cyan represents
        power-law, low-temperature plasma (vpshock), and high-temperature plasma (vnei) component.
        The red line of SW-A and SW-B are the gaussian indicated Fe-K${\beta}$ line.}
	\label{figure:allsightspctrum}
\end{figure}

We show the integrated spectra in each FoV in Fig.\ref{figure:allsightspctrum}. These spectra
have been corrected for NXB, using the spectra generated by xisnxbgen (\citealt{2008PASJ...60S..11T}).

\citet{2008PASJ...60S.123Y,2011PASJ...63S.837Y}
showed that the emission in RCW~86 can be well reproduced by
two temperature plasma and power-law component model, and taking into account Galactic absorption.
The low-temperature component corresponds to the shock-heated ISM, and is modeled
with the vpshock model. The high-temperature component coresponds to the heated ejecta and was
modeled by the vnei model. Finally, the power-law component represents the synchrotron emission from accelerated
electrons.
In vnei, we fixed the temperature and the ionization timescales to 5~keV
and $10^{9}$~cm$^{-3}$~s, respectively,
following \citet{2011PASJ...63S.837Y}.
Furthermore, the abundances were fixed to solar abundance except for iron,
for which we set the abundance to 10$^{10}$ solar
under the assumption that the ejecta plasma consists only of Fe ions and electrons
\citep{2011PASJ...63S.837Y}.
Because of uncertainties in the model, we ignored the energy bands of
0.70--0.78 and 1.13--1.28~keV (\citealt{2011PASJ...63S.837Y}). In addition, the 1.83--1.85~keV range
was ignored, as the instrument response is uncertain in this range.

Because of the differences of spectral shapes in SW (\citealt{2000PASJ...52.1157B},
\citealt{2001ApJ...550..334B}), we divided SW into SW-A and SW-B. SW-B contains
SW 03, 04, 12, and 13 (see Figure~\ref{figure:image}). The others were included in SW-A.
These regions are much brighter than other regions, thus we further included the following steps
to reduce uncertainty of the responses.
We applied a gaussian for Fe-K$\beta$ line
with the fixed central energy and the sigma of gaussian to 7.1~keV and 0~eV, respectively.
The gaussian smoothing ({\tt gsmooth} model in XSPEC),
with a freely varying sigma at 6~keV
and a fixed index of 0.5 to take account of the energy dependence,
is also applied
to mitigate residuals in the low-energy band likely due to calibration
uncertainty in instrumental broadening.
Furthermore, we treated the gain as a free parameter.

In the EAST,
there are the residuals of the NXB subtraction. When we checked the spectra from each FI
sensor, XIS~2 spectrum had more residuals than those of the other XISs,
implying that the low reproducibility
of NXB at XIS~2 causes these residuals. Therefore, we did not use the spectra of XIS~2 for the
analysis of EAST region. When we fitted EAST spectrum in 0.5--7.0~keV, the best-fit parameters
did not change significantly.

By the above-mentioned method, we obtained the reasonable fitting results
in the all of FoV for complicated SNR emission.
Table \ref{table:allsight} and \ref{table:SW1and2},
and Figure~\ref{figure:allsightspctrum} show the best-fit parameters and models.
Although it is not perfect in
some regions, the purpose of this paper is not to determine the thermal parameters but to measure
general trends in the  emission measure of the thermal emission and the properties of X-ray synchrotron emission.
Therefore, we do not pursue this model furthermore.

\subsubsection{Spectra in the divided regions}\label{regions}

In the next step, we analyzed the spectra of the divided regions.
The spectral parameters were fixed to those of each FoV region (Table~\ref{table:allsight} and \ref{table:SW1and2}),
except for the normalization of each component and the power-law indices which were treated as
free.
We chose the
background regions in each FoV, where there is no bright source nor a calibration source.
For the WEST region, we used the background for SE since the FoV of the WEST region only covers
regions with emission from the bright shell, without any obvious region from which to extract
a background spectrum.
We show the spectra and best-fit parameters
Tables
\ref{table:EAST_region}--\ref{table:WEST_region}, respectively. The model fits were generally good
given the small number of free parameters.

Our results can change with different AtomDB version.
We thus compared the fitting with v1.3.2, 2.0.2, and 3.0.3.
The largest difference is shown in the Fe L lines in the ISM component,
and it makes the difference of best-fit emission measure
with the factor of around 2.
All the other components did not show significant change among different
AtomDB versions.
Since our aim is to examine rough characteristics of thermal emission,
and the best-fit emission measure scatter to more than 4 orders of magnitude,
we concluded that the difference of AtomDB version does not change
our result significantly.

Table \ref{table:EAST_region} to \ref{table:WEST_region} show that SOUTH08 region has the hardest spectrum.
We also fitted the spectrum in this region with srcut model (\citealt{1999ApJ...525..368R})
instead of the power-law model addition to thermal emission, in order to estimate the roll-off
energy of synchrotron emission in this region. The index of srcut were fixed to 0.6
(\citealt{1988Ap&SS.148....3G}). As the $\chi^{2}$/dof is 147/155, the fit is acceptable, and
this model is slightly better than previous model with power-law.
We obtained the roll-off
energy $h\nu_{\rm{rolloff}} = 0.61^{+0.11}_{-0.19}$~keV.

\begin{figure}
\epsscale{0.8}
\plotone{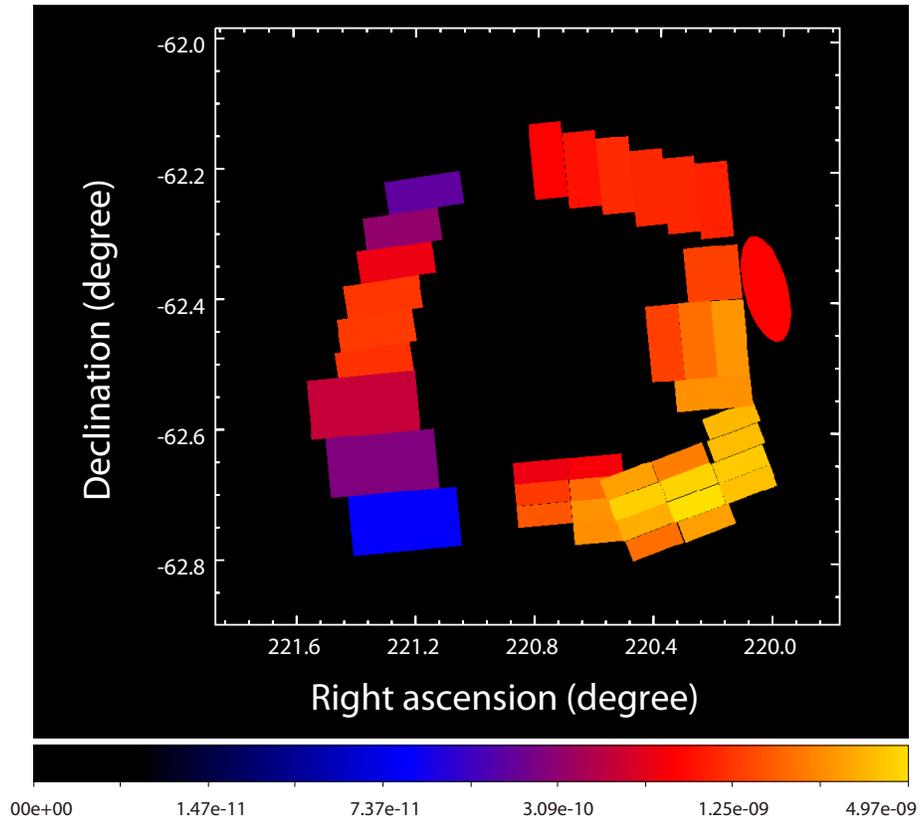}
\caption{%
The {\it EM}$^{1/2}$ map. The scale is logarithmic.
Note that SOUTH10 is invisible since this region has a very low $EM^{1/2}$ value
near the lower limit of the color map.
}
\label{figure:parmap-EM}
\end{figure}

\begin{figure}
\plotone{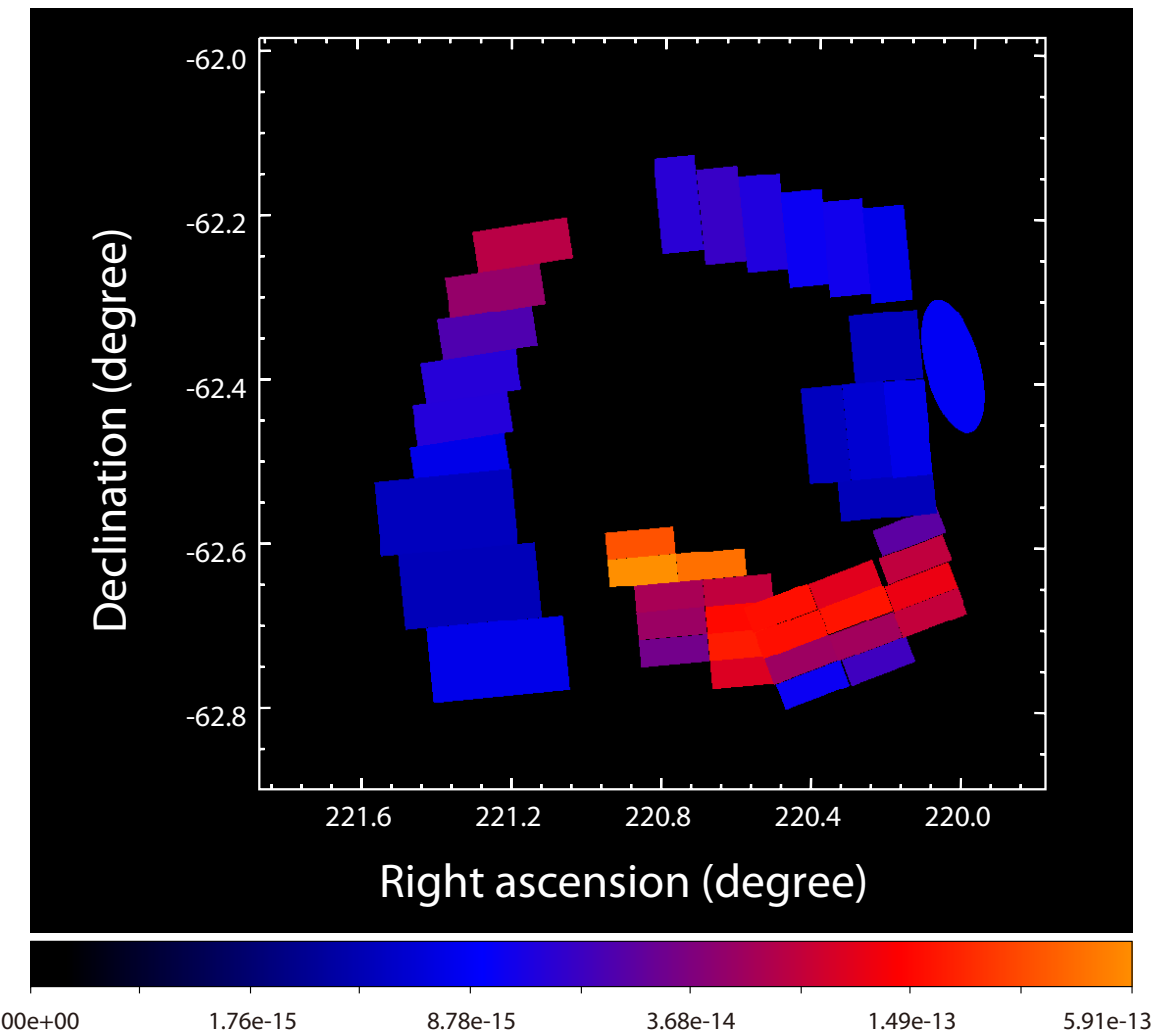}
\caption{%
The {\it SB}$_{\rm NT}$ map. The scale is lorarithmic.
}
\label{figure:parmap-SB}
\end{figure}

\begin{figure}
\plotone{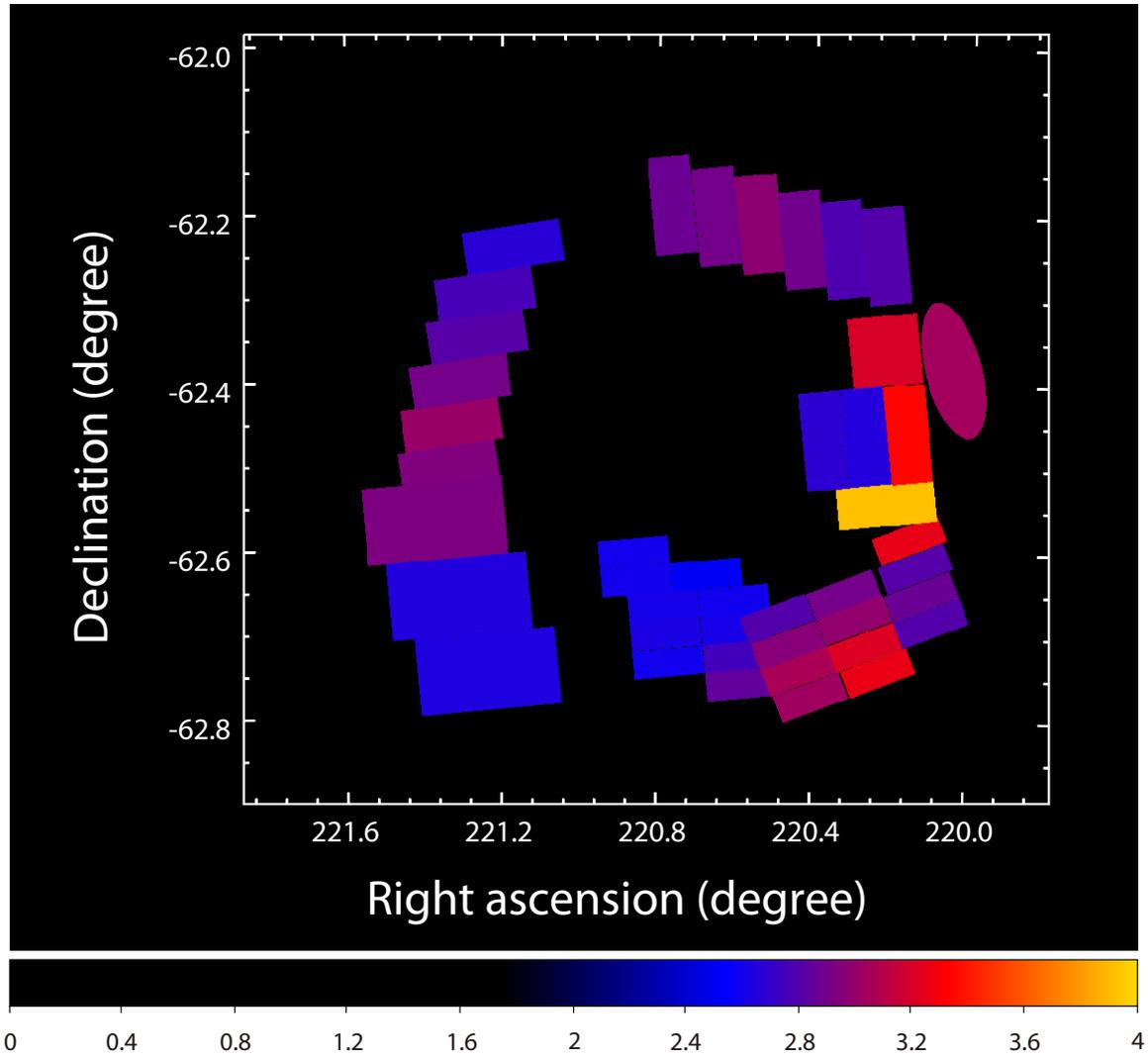}
\caption{The photon index map. The scale is linear.
}
\label{figure:parmap-index}
\end{figure}

In Figure~\ref{figure:parmap-EM}, \ref{figure:parmap-SB}, and \ref{figure:parmap-index},
we show the spatial
distributions of the the square root of the emission measure of the ISM heated plasma ({\it EM}$^{1/2}$),
the surface brightness of the synchrotron emission (the {\it SB}$_{\rm NT}$),
and the photon indices, respectively.
$EM^{1/2}$ is primarily proportional to the plasma density,
whereas {\it SB}$_{\rm NT}$ is proportional to a combination of 
the density of the accelerated electrons, the magnetic field strength,
and the maximum energy of the electrons in the region.
Finally, the photon index of synchrotron emission has correlation with the maximum
acceleration energy
and/or magnetic field strength
\citep[e.g.,][]{1999ApJ...525..368R}.
Toward southwest to west {\it EM}$^{1/2}$ and photon index become larger,
whereas {\it SB}$_{\rm NT}$ smaller.
In the northeast and
south regions, we have strong and hard synchrotron emission, and faint thermal emission.
%
%
According to Figure~\ref{figure:parmap-index}, we find the spatial difference of the photon
indices. Then we show the azimuthal dependence of photon indices in Figure~\ref{figure:azimuth}.
We set NORTH01 to 0$^{\circ}$. The emission goes hard around 190$^{\circ}$ (south),
and goes soft around 100$^{\circ}$ (southwest) and 260$^{\circ}$ (east).
We fit it with the constant model and rejected with
the $\chi^{2}$/dof of 672.4/42.
This result shows existence of azimuth angle dependence.

\section{Discussion}

We have conducted the systematic spatially-resolved spectroscopy of RCW~86,
and found that the spectral shape
changes from region to region.
Here, we discuss in which environment
the synchrotron emission becomes brighter and/or harder.

\begin{figure}
\epsscale{0.45}
\plotone{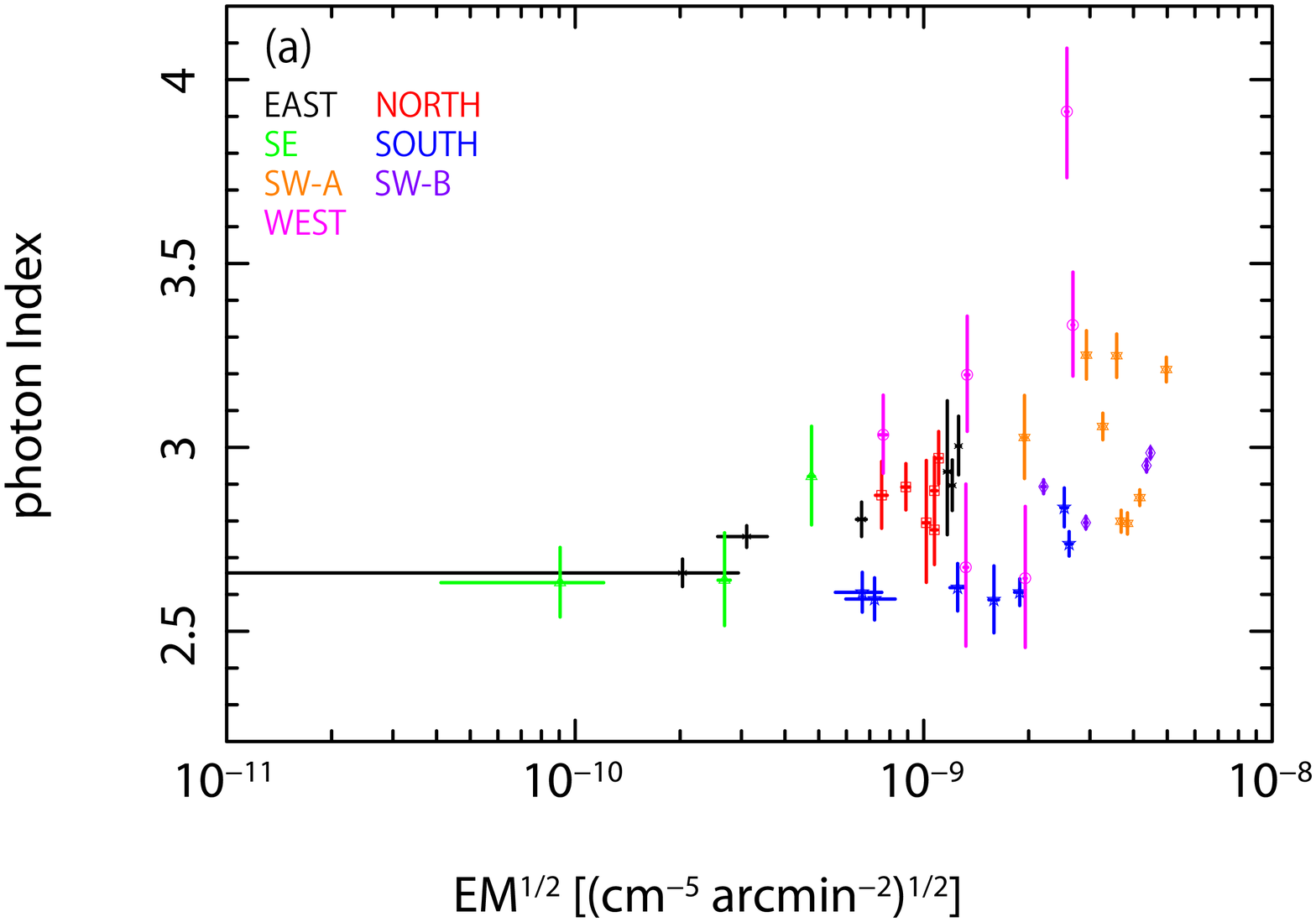}
\plotone{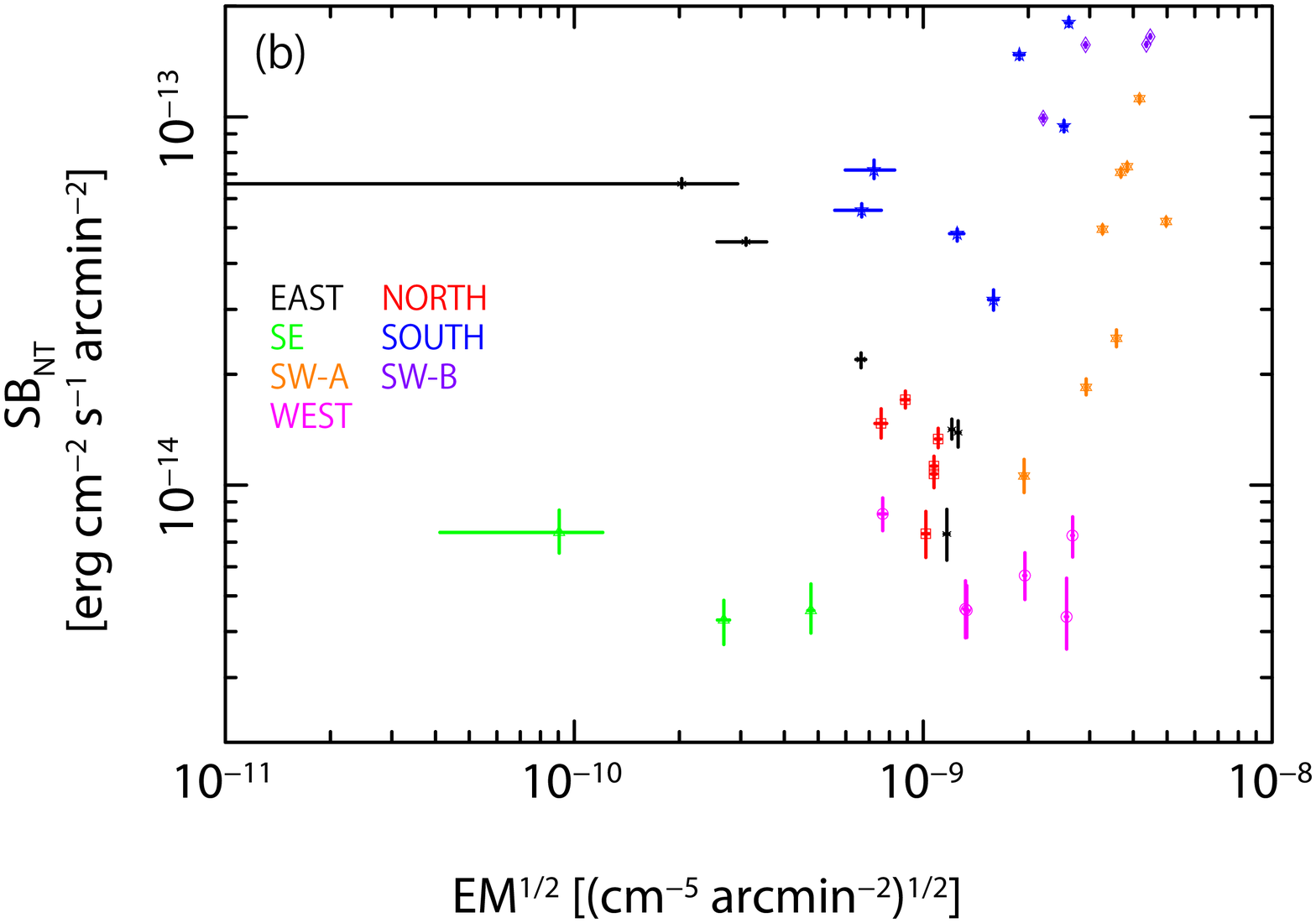}
	\caption{(a) Photon indices vs. EM$^{1/2}$,
          (b) SB$_{\rm{NT}}$ vs. EM$^{1/2}$.}
	\label{figure:allin}
\end{figure}

Figure~\ref{figure:allin} (a) and (b) show photon indices and {\it SB}$_{\rm NT}$ 
as a function of {\it EM}$^{1/2}$ (an indicator of the plasma density).
One can see that the photon index becomes smaller when
$EM^{1/2}$ becomes smaller.
The correlation coefficient is 0.38$\pm$0.15 (1$\sigma$ variance) $\pm$ 0.03 (1$\sigma$ statistical),
indicating a marginally positive correlation.
It implies that the acceleration to the higher energy in the low density medium.
The northeast and south regions are obvious cases with hard nonthermal emission and low ambient density,
whereas the high density region, such as southwest and west, shows softer nonthermal emission.
The northeast region is known to be an efficient acceleration site with high
shock velocity (\citealt{2013MNRAS.435..910H,yamaguchi2016}),
whereas the shock speed in southwest and north is much smaller
\citep{ghavamian2001,katsuda2014,fraschetti2016,long1990,2013ApJ...779...49C}.
So our results show that
the maximum energy of accelerated electrons is higher in high shock speed region.
This agrees with \citet{yamazaki2006} and \citet{aharonian1999},
which suggest that the roll-off of synchrotron emission depends only on the shock velocity
if the maximum electron energy is determined by the balance between accelerating and
synchrotron cooling.
In order for a more quantitative study of the dependence of the maximum energy on shock
velocity, we need more precise shock velocity measurement for the entire remnant.
The south region,
which is very interesting region with the hardest spectrum and very low ambient density in this remnant.
This region coinsides with the void of H$_{\rm I}$ cloud \citep{sano2016},
which is consistent with our scenario.

\begin{figure}
\epsscale{0.8}
\plotone{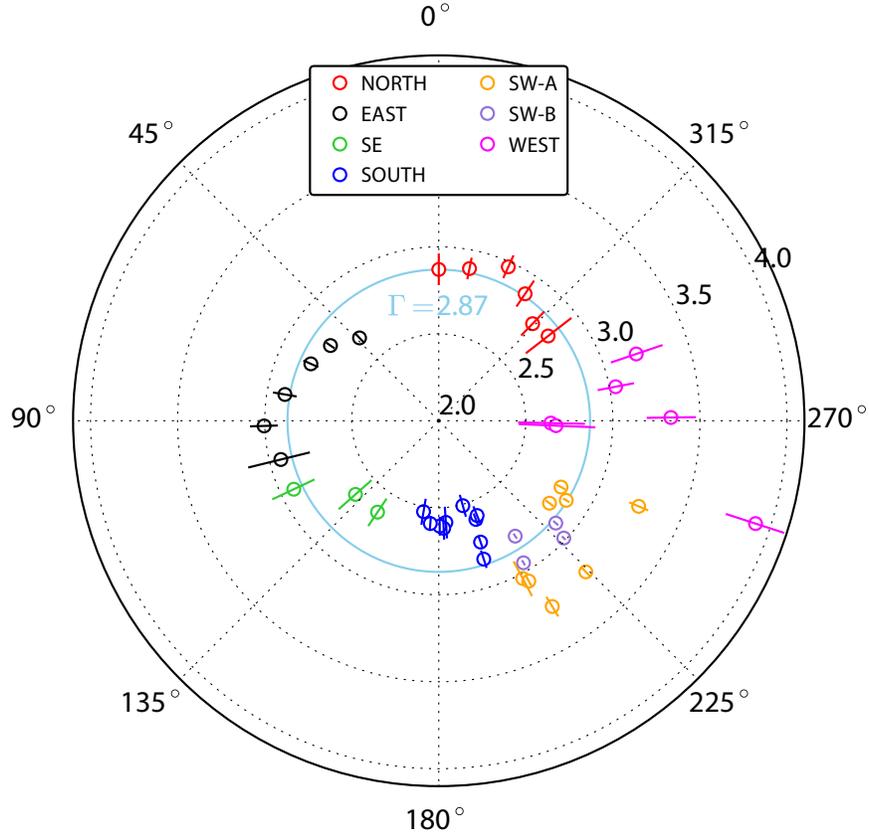}
	\caption{The azimuth angle dependence of photon indices. We set NORTH01 to 0$^{\circ}$.
	The cyan line shows the best-fit constant model.}
	\label{figure:azimuth}
\end{figure}

Figure~\ref{figure:azimuth} shows the complex azimthal dependence of photon index.
Azimuthal dependence is also reported in SN 1006
(\citealt{2004A&A...425..121R, 2009A&A...501..239M, 2010ApJ...723..383K}). SN 1006 has rather
simple bipolar morphology, and such a dependence is believed to be due to the effect of the magnetic
field parallel to the Galactic Plane (\citealt{2009MNRAS.393.1034P}) with the polarization
observations in the radio band (\citealt{2013AJ....145..104R}).
On the other hand, it is not rejected yet the effect of perpendicular magnetic field
(\citealt{2003ApJ...589..827B}), since Richtmyer-Meshkov instability can make radial magnetic
field in young SNRs even if the background magnetic field is perpendicular on the shock
(\citealt{2013ApJ...772L..20I}).
The situation is not so simple in our case. The photon
index is harder in the south region, but we have no bipolar trend.
This result indicates the reason of the photon index change may not the global magnetic field configuration
but the shock speed differences as shown before.
This may be because RCW 86 is evolving inside a stellar wind bubble \citep{2014MNRAS.441.3040B};
the stellar wind makes complicated shock evolution together with complex circumstellar medium structure.


On the other hand, we found no correlation between $EM^{1/2}$ and $SB_{\rm NT}$
as shown in Figure~\ref{figure:allin} (b),
which showed that we have no clear connection between the background plasma density
and the surface brightness of synchrotron emission.
The surface brightness of synchrotron emission depends not only on the number density of
accelerated particles but also the strength of magnetic field,
which makes perhaps for a more complicated relation between synchrotron surface brightness
and plasma density. 
It may also be that the magnetic field is turbulent on scales smaller 
than our region size.
It will be needed to compare these parameters in smaller size regions with a better angular resolution.

These results may be connected with the fact that many TeV gamma-ray SNRs have no significant thermal X-rays and
their plasma densities are believed to be very low. In practice, thermal emission is not found at
a significant level in Vela~Jr. (\citealt{2001ApJ...548..814S}), RX~J1713.7$-$3946
(\citealt{2008PASJ...60S.131T})\footnote{%
Recently, \citet{2015ApJ...814...29K} detected thermal X-rays with the ejecta origin
from the center of RX~J1713.7$-$3946, but we have no clue on interstellar
medium surrounding this SNR yet.}
,
and HESS~J1731-347 (\citealt{2012ApJ...756..149B}).
RCW 86 has shell-like TeV and GeV gamma-ray emission (\citealt{2009ApJ...692.1500A, 2014A&A...567A..23Y}),
in addition to the thermal and nonthermal X-rays.
Given that RCW 86 has regions with thermal emission and regions resembling the overall emission
characteristics of RX~J1713.7$-$3946 and Vela Jr, 
it may be a key object to understand the gamma-ray and
X-ray synchrotron emission of  these large TeV gamma-ray emitting shell-type  SNRs.
In the future, studies with the Cherenkov Telescope Array (CTA) (\citealt{2011ExA....32..193A,acharya2013}) will reveal us the
detailed morphology of RCW 86 with better spatial resolution and sensitivity, allowing for more exploring in more detail correlations
between the thermal and non-thermal X-ray emission and the  gamma-ray emission.

\acknowledgments

We thank the anonymous referee for the fruitful comments.
We also would like to thank R. Yamazaki, Y. Ohira, and J. Shimoda for their fruitful discussion.
This work is supported in part by
Grant-in-Aid for Scientific Research of
the Japanese Ministry of Education, Culture, Sports,
Science and Technology (MEXT) of Japan,
No.~22684012 and 15K05107 (A.~B.), 15K17657 (M.~S.), and 16K17673 (S.~K.).

{\it Facilities:}
\facility{Suzaku}

\begin{figure}
\epsscale{0.5}
\plotone{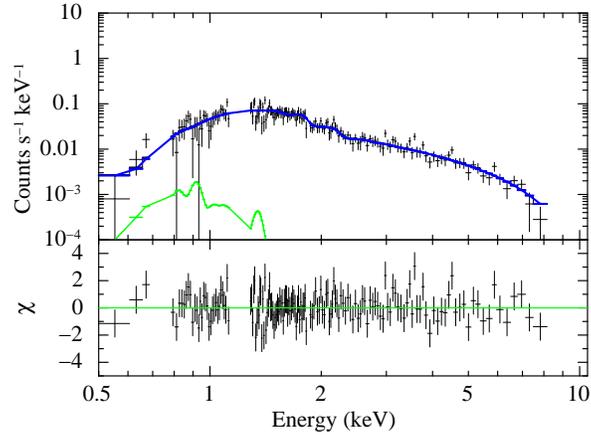}
	\caption{
	The SOUTH08 spectrum with with srcut fixed photon index and freed radio flux.
	The color is equivalent to Figure~\ref{figure:allsightspctrum}.}
	\label{figure:radiofix}
\end{figure}

\begin{deluxetable}{lccccccc}
\tabletypesize{\scriptsize}
\tablewidth{0pt}
\tablecolumns{8}
\tablecaption{Suzaku observations of RCW~86 \label{table:data}}
\tablehead{
	\colhead{Region} & \colhead{Observation ID} & \colhead{Start date} & \colhead{Exposure} & \colhead{R.A.} & \colhead{Decl.} & \colhead{\# of XIS} & \colhead{SCI}\\
	\colhead{} & \colhead{}& \colhead{yyyy/mm/dd}  & \colhead{(ks)} & \colhead{(deg)} & \colhead{(deg)} & \colhead{} & \colhead{} 
}
\startdata
	SW & 500004010 & 2006/02/12 & 100.8 & 220.28 & $-$62.68 & 4 & off \\ 
	EAST & 501037010 & 2006/08/12 & 59.8 & 221.26 & $-$62.36 & 4 & off \\ 
	NORTH & 503002010 & 2009/01/29 & 55.4 & 220.50 & $-$62.21 & 3 & on \\ 
	SOUTH & 503003010 & 2009/01/31 & 54.8 & 220.83 & $-$62.67 & 3 & on \\ 
	SE & 503004010 & 2009/02/01 & 53.5 & 221.39 & $-$62.67 & 3 & on \\ 
	WEST &503001010 & 2009/02/02 & 53.6 & 220.28 & $-$62.43 & 3 & on
\enddata
\end{deluxetable}

\begin{deluxetable}{lccccc}
\tabletypesize{\scriptsize}
\tablewidth{0pt}
\tablecolumns{6}
\tablecaption{The best-fit parameters for the each FoV. \label{table:allsight}}
\tablehead{\colhead{Parameter} & \colhead{EAST} & \colhead{NORTH} & \colhead{SE} & \colhead{SOUTH} & \colhead{WEST}
	}
\startdata
	\multicolumn{2}{l}{--wabs--} & & & & \\
	\hspace{3mm}$N_{\rm{H}}$ (10$^{21} $cm$^{-2}$) & 3.3$\pm$0.1 & 3.4$\pm$0.1 & 2.1$\pm$0.1 & 4.2$\pm$0.1 & 3.3$\pm$0.1 \\
	\multicolumn{2}{l}{--vpshock--} & & & & \\
	\hspace{3mm}$kT$ (keV) & 0.44$\pm$0.03 & 0.46$\pm$0.02 & 0.59$\pm$0.07 & 0.36$\pm$0.02 & 0.46$\pm$0.01 \\
	\hspace{3mm}Ne (solar) & 1.9$\pm$0.1 & 1.8$\pm$0.1 & 2.4$_{-0.1}^{+0.2}$ & 1.4$\pm$0.1 & 1.8$\pm$0.1 \\
	\hspace{3mm}Mg (solar) & 1.9$\pm$0.2 & 1.3$\pm$0.1 & 2.0$_{-0.3}^{+0.4}$ & 0.96$_{-0.12}^{+0.16}$ & 1.0$\pm$0.1 \\
	\hspace{3mm}Si (solar) & 1.9$_{-0.5}^{+0.6}$ & 1.8$_{-0.3}^{+0.4}$ & 3.2$_{-0.9}^{+1.3}$ & 1.2$_{-0.4}^{+0.5}$ & 1.4$\pm$0.1 \\
	\hspace{3mm}Ca (solar) & 1.2$\pm$0.1 & 0.73$_{-0.06}^{+0.07}$ & 1.1$_{-0.1}^{+0.2}$ & 0.93$_{-0.07}^{+0.09}$ & 0.62$\pm$0.02 \\
	\hspace{3mm}$\tau$ (10$^{10}$ s cm$^{-3}$) & 3.2$_{-0.5}^{+0.9}$ & 3.2$_{-0.5}^{+0.6}$ & 2.5$_{-0.5}^{+1.0}$ & 9.1$_{-1.6}^{+2.6}$ & 5.7$\pm$0.4 \\
	\hspace{3mm}Norm (10$^{-16}$ cm$^{-5}$) & 1.14$\pm$0.02 & 1.30$\pm$0.02 & 0.218$\pm$0.005 & 2.91$\pm$0.04 & 4.28$\pm$0.03 \\
	\multicolumn{2}{l}{--power-law--} & & & & \\
	\hspace{3mm}$\Gamma$ & 2.72$\pm$0.02 & 2.81$\pm$0.02 & 2.49$\pm$0.03 & 2.70$\pm$0.01 & 2.86$\pm$0.03 \\
	\hspace{3mm}Norm\tablenotemark{a} & 1.30$\pm$0.02 & 0.84$\pm$0.02 & 0.43$\pm0.01$ & 2.74$\pm$0.03 & 0.70$\pm0.02$ \\
	\multicolumn{2}{l}{--vnei--} & & & & \\
	\hspace{3mm}Norm (10$^{-21}$ cm$^{-5}$) & 4.6$\pm$0.8 & 4.5$\pm$0.7 & 2.9$\pm$0.7 & 5.6$\pm$1.0 & 6.9$\pm$0.8 \\ \tableline
	$\chi^{2}/dof$ & 550/464 & 385/314 & 382/349 & 745/583 & 576/323
\enddata
\tablenotetext{a}{In unit of 10$^{-2}$ photons keV$^{-1}$ s$^{-1}$ cm$^{-2}$ at 1~keV.}
\end{deluxetable}

\begin{deluxetable}{lcc}
\tabletypesize{\scriptsize}
\tablewidth{0pt}
\tablecolumns{3}
\tablecaption{The best-fit parameters for the SW regions. \label{table:SW1and2}}
\tablehead{\colhead{Parameter} & \colhead{SW-A} & \colhead{SW-B} }
\startdata
	\multicolumn{2}{l}{--gsmooth--} & \\
	\hspace{3mm}$\sigma$ (eV at 6~keV) & 24$\pm$1 & 30$\pm$2 \\
	\multicolumn{2}{l}{--wabs--} & \\
	\hspace{3mm}$N_{\rm{H}}$ (10$^{21} $cm$^{-2}$) & 3.7$\pm$0.1 & 4.5$\pm$0.1 \\
	\multicolumn{2}{l}{--vpshock--} & \\
	\hspace{3mm}$kT$ (keV) & 0.438$_{-0.002}^{+0.004}$ & 0.35$\pm$0.01 \\
	\hspace{3mm}Ne (solar) & 1.85$_{-0.02}^{+0.01}$ & 1.49$\pm$0.05 \\
	\hspace{3mm}Mg (solar) & 1.66$_{-0.03}^{+0.02}$ & 1.01$\pm$0.07 \\
	\hspace{3mm}Si (solar) & 1.62$_{-0.05}^{+0.04}$ & 1.15$_{-0.18}^{+0.20}$ \\
	\hspace{3mm}Ca (solar) & 1.02$\pm$0.01 & 0.83$\pm0.03$ \\
	\hspace{3mm}$\tau$ (10$^{10}$ s cm$^{-3}$) & 6.6$_{-0.2}^{+0.1}$ & 8.3$_{-0.8}^{+0.9}$ \\
	\hspace{3mm}Norm (10$^{-16}$ cm$^{-5}$) & 8.36$\pm$0.02 & 4.21$\pm$0.03 \\
	\multicolumn{2}{l}{--power-law--} & \\
	\hspace{3mm}$\Gamma$ & 3.06$\pm$0.01 & 2.91$\pm$0.01 \\
	\hspace{3mm}Norm\tablenotemark{a} & 1.60$\pm$0.02 & 1.85$\pm$0.01 \\
	\multicolumn{2}{l}{--vnei--} & \\
	\hspace{3mm}Norm (10$^{-21}$ cm$^{-5}$) & 8.9$\pm$0.5 & 5.9$\pm$0.4 \\
	\multicolumn{2}{l}{--gaussian--} & \\
	\hspace{3mm}Norm\tablenotemark{b} & 2.8$\pm$1.2 & 1.1$_{-1.1}^{+1.2}$ \\
	\multicolumn{2}{l}{--gain--} & \\
	\hspace{3mm}slope & \multicolumn{2}{c}{1.0032$\pm$0.0001} \\
	\hspace{3mm}offset (eV) & \multicolumn{2}{c}{-1.5$\pm$0.1} \\ \tableline
	$\chi^{2}/dof$ & 2050/905 & 1557/1026
\enddata
\tablenotetext{a}{In unit of 10$^{-2}$ photons keV$^{-1}$ s$^{-1}$ cm$^{-2}$ at 1~keV.}
\tablenotetext{b}{In unit of 10$^{-6}$ photons s$^{-1}$ cm$^{-2}$ in the line.}
\end{deluxetable}

\clearpage
\begin{deluxetable}{lcccccc}
\tabletypesize{\scriptsize}
\tablewidth{0pt}
\tablecolumns{7}
\tablecaption{The best-fit parameter for all EAST regions \label{table:EAST_region}}
\tablehead{\colhead{Parameter} & \colhead{EAST01} & \colhead{EAST02} & \colhead{EAST03} & \colhead{EAST04} & \colhead{EAST05} & \colhead{EAST06}}
\startdata
	\multicolumn{2}{l}{--vpshock--} & & & & & \\
	\hspace{3mm}Norm\tablenotemark{a} & 2.9$\pm$0.1 & 3.3$\pm$0.1 & 3.1$\pm$0.1 & 0.93$\pm$0.06 & 0.20$\pm$0.06 & 0.06$_{-0.06}^{+0.07}$ \\
	\multicolumn{2}{l}{--Power-law--} & & & & & \\
	\hspace{3mm}$\Gamma$ & 2.9$\pm$0.2 & 3.00$\pm$0.08 & 2.90$\pm$0.07 & 2.80$\pm$0.05 & 2.76$\pm$0.03 & 2.66$\pm$0.04 \\
	\hspace{3mm}Norm\tablenotemark{b} & 0.7$\pm 0.1$ & 1.4$\pm$0.1 & 1.2$\pm$0.1 & 1.7$\pm$0.1 & 3.4$\pm$0.1 & 3.1$\pm$0.1 \\
	\hspace{3mm}F$_X$\tablenotemark{c} & 1.6$_{-0.2}^{+0.3}$ & 3.0$_{-0.3}^{+0.2}$ & 3.0$\pm$0.2 & 4.6$\pm$0.2 & 9.6$\pm$0.2 & 10.1$_{-0.2}^{+0.3}$ \\
	\multicolumn{2}{l}{--vnei--} & & & & & \\
	\hspace{3mm}Norm\tablenotemark{d} & $<$9.9 & 9.8$\pm$4.2 & 7.5$\pm$2.2 & $<$4.4 & $<$0.9 & $<$0.9 \\ \tableline
	$\chi^{2}/dof$ & 219/184 & 304/271 & 330/285 & 357/290 & 468/399 & 265/250 
\enddata
\tablenotetext{a}{In unit of 10$^{-17}$ cm$^{-5}$.}
\tablenotetext{b}{In unit of 10$^{-3}$ photons keV$^{-1}$ s$^{-1}$ cm$^{-2}$ at 1~keV.}
\tablenotetext{c}{In unit of 10$^{-13}$ erg s$^{-1}$ cm$^{-2}$ in the 3.0--5.0~keV band.}
\tablenotetext{d}{In unit of 10$^{-22}$ cm$^{-5}$.}
\end{deluxetable}

\begin{deluxetable}{lcccccc}
\tabletypesize{\scriptsize}
\tablewidth{0pt}
\tablecolumns{7}
\tablecaption{The best-fit parameter for all NORTH regions \label{table:NORTH_region}}
\tablehead{\colhead{Parameter} & \colhead{NORTH01} & \colhead{NORTH02} & \colhead{NORTH03} & \colhead{NORTH04} & \colhead{NORTH05} & \colhead{NORTH06}}
\startdata
	\multicolumn{2}{l}{--vpshock--} & & & & & \\
	\hspace{3mm}Norm\tablenotemark{a} & 1.1$\pm$0.1 & 1.7$\pm$0.1 & 2.6$\pm$0.1 & 2.4$\pm$0.1 & 2.4$\pm$0.1 & 2.2$\pm$0.1 \\
	\multicolumn{2}{l}{--power-law--} & & & & & \\
	\hspace{3mm}$\Gamma$ & 2.87$\pm$0.09 & 2.89$\pm$0.06 & 2.97$\pm$0.07 & 2.88$\pm$0.09 & 2.78$_{-0.09}^{+0.10}$ & 2.8$\pm$0.2 \\
	\hspace{3mm}Norm\tablenotemark{b} & 1.1$\pm$0.1 & 1.5$\pm$0.1 & 1.3$\pm$0.1 & 0.9$\pm$0.1 & 0.9$\pm$0.1 & 0.6$\pm$0.1 \\
	\hspace{3mm}F$_X$\tablenotemark{c} & 2.8$_{-0.2}^{+0.3}$ & 3.6$\pm$0.2 & 2.8$\pm$0.2 & 2.3$\pm$0.2 & 2.4$\pm$0.2 & 1.6$\pm$0.2 \\
	\multicolumn{2}{l}{--vnei--} & & & & & \\
	\hspace{3mm}Norm\tablenotemark{d} & $<$6.2 & 5.1$\pm$3.7 & 3.4$\pm$2.8 & 6.1$\pm$3.3 & 8.9$\pm$4.2 & $<$10.3 \\ \tableline
	$\chi^{2}/dof$ & 163/133 & 226/200 & 200/226 & 200/190 & 169/164 & 160/124
\enddata
\tablenotetext{a}{In unit of 10$^{-17}$ cm$^{-5}$.}
\tablenotetext{b}{In unit of 10$^{-3}$ photons keV$^{-1}$ s$^{-1}$ cm$^{-2}$ at 1~keV.}
\tablenotetext{c}{In unit of 10$^{-13}$ erg s$^{-1}$ cm$^{-2}$ in the 3.0--5.0~keV band.}
\tablenotetext{d}{In unit of 10$^{-22}$ cm$^{-5}$.}
\end{deluxetable}

\begin{deluxetable}{lccc}
\tabletypesize{\scriptsize}
\tablewidth{0pt}
\tablecolumns{4}
\tablecaption{The best-fit parameter for all SE regions \label{table:SE_region}}
\tablehead{\colhead{Parameter} & \colhead{SE01} & \colhead{SE02} & \colhead{SE03}}
\startdata
	\multicolumn{2}{l}{--vpshock--} & & \\
	\hspace{3mm}Norm\tablenotemark{a} & 1.3$\pm$0.1 & 0.40$\pm$0.03 & 0.05$\pm$0.04 \\
	\multicolumn{2}{l}{--power-law--} & & \\
	\hspace{3mm}$\Gamma$ & 2.92$_{-0.13}^{+0.14}$ & 2.64$_{-0.12}^{+0.13}$ & 2.63$_{-0.09}^{+0.10}$ \\
	\hspace{3mm}Norm\tablenotemark{b} & 1.1$\pm$0.1 & 0.7$\pm$0.1 & 1.2$\pm$0.1 \\
	\hspace{3mm}F$_X$\tablenotemark{c} & 2.6$_{-0.3}^{+0.5}$ & 2.4$\pm$0.3 & 4.1$_{-0.5}^{+0.6}$ \\
	\multicolumn{2}{l}{--vnei--} & & \\
	\hspace{3mm}Norm\tablenotemark{d} & 8.4$\pm$6.1 & 8.0$\pm$5.1 & $<$12.2 \\ \tableline
	$\chi^{2}/dof$ & 298/259 & 220/197 & 204/186
\enddata
\tablenotetext{a}{In unit of 10$^{-17}$ cm$^{-5}$.}
\tablenotetext{b}{In unit of 10$^{-3}$ photons keV$^{-1}$ s$^{-1}$ cm$^{-2}$ at 1~keV.}
\tablenotetext{c}{In unit of 10$^{-13}$ erg s$^{-1}$ cm$^{-2}$ in the 3.0--5.0~keV band.}
\tablenotetext{d}{In unit of 10$^{-22}$ cm$^{-5}$.}
\end{deluxetable}

\begin{deluxetable}{lccccc}
\tabletypesize{\scriptsize}
\tablewidth{0pt}
\tablecolumns{11}
\tablecaption{The best-fit parameter for all SOUTH regions \label{table:SOUTH_region}}
\tablehead{\colhead{Parameter} & \colhead{SOUTH01} & \colhead{SOUTH02} & \colhead{SOUTH03} & \colhead{SOUTH04} & \colhead{SOUTH05}}
\startdata
	\multicolumn{2}{l}{--vpshock--} \\
	\hspace{3mm}Norm\tablenotemark{a} & 6.4$\pm$0.3 & 6.8$\pm$0.3 & 3.6$\pm$0.2 & 0.5$\pm$0.2 & 2.5$\pm$0.2 \\
	\multicolumn{2}{l}{--power-law--} \\
	\hspace{3mm}$\Gamma$ & 2.84$\pm$0.05 & 2.74$\pm$0.03 & 2.61$\pm$0.04 & 2.59$\pm$0.06 & 2.59$\pm$0.09 \\
	\hspace{3mm}Norm\tablenotemark{b} & 3.7$\pm$0.2 & 6.2$\pm$0.2 & 4.2$\pm$0.1 & 2.0$\pm$0.1 & 0.9$\pm$0.1 \\
	\hspace{3mm}F$_X$\tablenotemark{c} & 9.5$\pm$0.3 & 18.1$_{-0.4}^{+0.6}$ & 14.8$_{-0.4}^{+0.3}$ & 7.2$\pm$0.4 & 3.2$\pm$0.2 \\
	\multicolumn{2}{l}{--vnei--} \\
	\hspace{3mm}Norm\tablenotemark{d} & $<$2.0 & $<$7.7 & $<$5.1 & 6.1$\pm$5.6 & $<$2.1 \\
	$\chi^{2}/dof$ & 298/288 & 445/413 & 373/352 & 199/199 & 195/175 \\ \tableline\tableline
 & SOUTH06 & SOUTH07 & SOUTH08 & SOUTH09 & SOUTH10 \\ \tableline
	\multicolumn{2}{l}{--vpshock--} \\
	\hspace{3mm}Norm\tablenotemark{a} & 1.6$\pm$0.1 & 0.4$\pm$0.1 & $<$0.05 & $<$0.04 & $<$0.02 \\
	\multicolumn{2}{l}{--power-law--}  \\
	\hspace{3mm}$\Gamma$ & 2.62$_{-0.06}^{+0.07}$ & 2.61$\pm$0.05 & 2.51$_{-0.06}^{+0.07}$ & 2.59$\pm$0.04 & 2.53$\pm$0.08 \\
	\hspace{3mm}Norm\tablenotemark{b} & 1.4$\pm$0.1 & 1.6$\pm$0.1 & 1.1$\pm$0.1 & 1.7$\pm$0.1 & 0.8$\pm$0.1 \\
	\hspace{3mm}F$_X$\tablenotemark{c} & 4.8$_{-0.2}^{+0.1}$ & 5.6$\pm$0.2 & 4.3$_{-0.3}^{+0.2}$ & 5.9$\pm$0.2 & 3.2$\pm$0.2 \\
	\multicolumn{2}{l}{--vnei--} \\
	\hspace{3mm}Norm\tablenotemark{d} & $<$3.5 & 3.8$\pm$3.7 & $<$0.9 & $<$1.6 & $<$1.4 \\
	$\chi^{2}/dof$ & 207/203 & 208/213 & 153/156 & 206/215 & 148/140 \\ 
\enddata
\tablenotetext{a}{In unit of 10$^{-17}$ cm$^{-5}$.}
\tablenotetext{b}{In unit of 10$^{-3}$ photons keV$^{-1}$ s$^{-1}$ cm$^{-2}$ at 1~keV.}
\tablenotetext{c}{In unit of 10$^{-13}$ erg s$^{-1}$ cm$^{-2}$ in the 3.0--5.0~keV band.}
\tablenotetext{d}{In unit of 10$^{-22}$ cm$^{-5}$.}
\end{deluxetable}

\begin{deluxetable}{lcccccccc}
\tabletypesize{\scriptsize}
\tablewidth{0pt}
\tablecolumns{9}
\tablecaption{The best-fit parameter for all SW-A regions \label{table:SW1_region}}
\tablehead{\colhead{Parameter} & \colhead{SW01} & \colhead{SW02} & \colhead{SW05} & \colhead{SW06} & \colhead{SW07} & \colhead{SW08} & \colhead{SW09} & \colhead{SW11}}
\startdata
	\multicolumn{2}{l}{--vpshock--} & & & & & & & \\
	\hspace{3mm}Norm\tablenotemark{a} & 3.8$\pm$0.1 & 10.7$\pm$0.1 & 8.6$\pm$0.1 & 14.8$\pm$0.1 & 17.4$\pm$0.1 & 13.6$\pm$0.1 & 12.8$\pm$0.1 & 24.7$\pm$0.1 \\
	\multicolumn{2}{l}{--power-law--} & & & & & & & \\
	\hspace{3mm}$\Gamma$ & 3.03$\pm$0.11 & 3.06$\pm$0.04 & 3.25$\pm$0.07 & 2.79$\pm$0.03 & 2.86$\pm$0.02 & 2.80$\pm$0.03 & 3.25$\pm$0.06 & 3.21$\pm$0.03 \\
	\hspace{3mm}Norm\tablenotemark{b} & 0.5$\pm$0.1 & 2.6$\pm$0.1 & 1.2$\pm$0.1 & 2.7$\pm$0.1 & 4.5$\pm$0.1 & 2.6$\pm$0.1 & 1.7$\pm$0.1 & 3.3$\pm$0.1 \\
	\hspace{3mm}F$_X$\tablenotemark{c} & 1.1$\pm$0.1 & 5.1$\pm$0.1 & 1.9$\pm$0.1 & 7.5$\pm$0.2 & 11.4$\pm$0.2 & 7.2$\pm$0.2 & 2.6$\pm$0.1 & 5.4$\pm$0.1 \\
	\multicolumn{2}{l}{--vnei--} & & & & & & & \\
	\hspace{3mm}Norm\tablenotemark{d} & 4.5$\pm$1.8 & 11$\pm$0.2 & 9.8$\pm$1.9 & 14$\pm$2.3 & 21$\pm$2.8 & 13$\pm$2.3 & 8.0$\pm$1.9 & 29$\pm$2.8 \\
	\multicolumn{2}{l}{--gaussian--} & & & & & & & \\
	\hspace{3mm}Norm\tablenotemark{e} & $<$11.1 & $<$12.0 & $<$8.4 & $<$3.8 & $<$4.6 & $<$4.9 & $<$8.1 & 13$\pm$6 \\ \tableline
	$\chi^{2}/dof$ & 337/280 & 762/498 & 978/408 & 1317/572 & 1227/690 & 986/560 & 764/424 & 1460/585
\enddata
\tablenotetext{a}{In unit of 10$^{-17}$ cm$^{-5}$}
\tablenotetext{b}{In unit of 10$^{-3}$ photons keV$^{-1}$ s$^{-1}$ cm$^{-2}$ at 1~keV.}
\tablenotetext{c}{In unit of 10$^{-13}$ erg s$^{-1}$ cm$^{-2}$ in the 3.0--5.0~keV band.}
\tablenotetext{d}{In unit of 10$^{-22}$ cm$^{-5}$.}
\tablenotetext{e}{In unit of 10$^{-7}$ photons s$^{-1}$ cm$^{-2}$ in the line.}
\end{deluxetable}

\begin{deluxetable}{lcccc}
\tabletypesize{\scriptsize}
\tablewidth{0pt}
\tablecolumns{5}
\tablecaption{The best-fit parameter for all SW-B regions \label{table:SW2_region}}
\tablehead{\colhead{Parameter} & \colhead{SW03} & \colhead{SW04} & \colhead{SW12} & \colhead{SW13}}
\startdata
	\multicolumn{2}{l}{--vpshock--} & & & \\
	\hspace{3mm}Norm\tablenotemark{a} & 19.0$\pm$0.2 & 8.6$\pm$0.2 & 20.1$\pm$0.2 & 4.9$\pm$0.1 \\
	\multicolumn{2}{l}{--power-law--} & & & \\
	\hspace{3mm}$\Gamma$ & 2.95$\pm$0.02 & 2.80$\pm$0.02 & 2.98$\pm$0.01 & 2.89$\pm$0.02 \\
	\hspace{3mm}Norm\tablenotemark{b} & 7.2$\pm$0.1 & 5.8$\pm$0.1 & 7.9$\pm$0.1 & 4.2$\pm$0.1 \\
	\hspace{3mm}F$_X$\tablenotemark{c} & 15.8$\pm$0.2 & 15.7$\pm$0.2 & 16.6$\pm$0.2 & 10.0$\pm$0.2 \\
	\multicolumn{2}{l}{--vnei--} & & & \\
	\hspace{3mm}Norm\tablenotemark{d} & 17$\pm$3.3 & 4.5$\pm$2.8 & 30$\pm$3.3 & 22$\pm$2.8 \\
	\multicolumn{2}{l}{--gaussian--} & & & \\
	\hspace{3mm}Norm\tablenotemark{e} & $<$12.7 & $<$4.7 & 11$\pm$8 & $<$11.8 \\ \tableline
	$\chi^{2}/dof$ & 917/755 & 887/716 & 1059/869 & 768/677
\enddata
\tablenotetext{a}{In unit of 10$^{-17}$ cm$^{-5}$.}
\tablenotetext{b}{In unit of 10$^{-3}$ photons keV$^{-1}$ s$^{-1}$ cm$^{-2}$ at 1~keV.}
\tablenotetext{c}{In unit of 10$^{-13}$ erg s$^{-1}$ cm$^{-2}$ in the 3.0--5.0~keV band.}
\tablenotetext{d}{In unit of 10$^{-22}$ cm$^{-5}$.}
\tablenotetext{e}{In unit of 10$^{-7}$ photons s$^{-1}$ cm$^{-2}$ in the line.}
\end{deluxetable}

\begin{deluxetable}{lcccccc}
\tabletypesize{\scriptsize}
\tablewidth{0pt}
\tablecolumns{7}
\tablecaption{The best-fit parameter for all WEST regions \label{table:WEST_region}}
\tablehead{\colhead{Parameter} & \colhead{WEST01} & \colhead{WEST02} & \colhead{WEST03} & \colhead{WEST04} & \colhead{WEST05} & \colhead{WEST06}}
\startdata
	\multicolumn{2}{l}{--vpshock--} & & & & & \\
	\hspace{3mm}Norm\tablenotemark{a} & 14.0$\pm$0.2 & 15.1$\pm$0.2 & 8.0$\pm$0.1 & 3.7$\pm$0.1 & 4.5$\pm$0.1 & 1.8$\pm$0.1 \\
	\multicolumn{2}{l}{--power-law--} & & & & & \\
	\hspace{3mm}$\Gamma$ & 3.9$\pm$0.2 & 3.3$\pm$0.1 & 2.6$\pm$0.2 & 2.7$\pm$0.2 & 3.2$\pm$0.2 & 3.0$\pm$0.1 \\
	\hspace{3mm}Norm\tablenotemark{b} & 1.5$\pm$0.2 & 1.2$\pm$0.1 & 0.4$\pm$0.1 & 0.3$\pm$0.1 & 0.7$\pm$0.1 & 1.3$\pm$0.1 \\
	\hspace{3mm}F$_X$\tablenotemark{c} & 1.1$_{-0.2}^{+0.3}$ & 1.7$\pm$0.2 & 1.3$\pm$0.2 & 1.0$\pm$0.2 & 1.2$\pm$0.2 & 2.6$\pm$0.3\\
	\multicolumn{2}{l}{--vnei--} & & & & & \\
	\hspace{3mm}Norm\tablenotemark{d} & 13$\pm$5.5 & 8.0$\pm$3.7 & 12$\pm$3.3 & 13$\pm$3.7 & 8.9$\pm$3.3 & $<$6.1 \\ \tableline
	$\chi^{2}/dof$ & 510/243 & 452/281 & 307/233 & 205/164 & 275/208 & 231/182
\enddata
\tablenotetext{a}{In unit of 10$^{-17}$ cm$^{-5}$.}
\tablenotetext{b}{In unit of 10$^{-3}$ photons keV$^{-1}$ s$^{-1}$ cm$^{-2}$ at 1~keV.}
\tablenotetext{c}{In unit of 10$^{-13}$ erg s$^{-1}$ cm$^{-2}$ in the 3.0--5.0~keV band.}
\tablenotetext{d}{In unit of 10$^{-22}$ cm$^{-5}$.}
\end{deluxetable}


\begin{thebibliography}{99}
\bibitem[Acharya et al.(2013)]{acharya2013}
Acharya, B.~S., Actis, M., Aghajani, T., et al.\ 2013, Astroparticle Physics, 43, 3 
\bibitem[Ackermann et al.(2013)]{2013Sci...339..807A}
 Ackermann, M., Ajello, M., Allafort, A., et al.\ 2013, Science, 339, 807 
\bibitem[Actis et al.(2011)]{2011ExA....32..193A}
 Actis, M., Agnetta, G., Aharonian, F., et al.\ 2011, Experimental Astronomy, 32, 193
\bibitem[Aharonian \& Atoyan(1999)]{aharonian1999}
Aharonian, F.~A., \& Atoyan, A.~M.\ 1999, \aap, 351, 330 
\bibitem[Aharonian et al.(2009)]{2009ApJ...692.1500A}
 Aharonian, F., Akhperjanian, A.~G., de Almeida, U.~B., et al.\ 2009, \apj, 692, 1500 
\bibitem[Bamba et al.(2000)]{2000PASJ...52.1157B}
 Bamba, A., Koyama, K., \& Tomida, H.\ 2000, PASJ, 52, 1157 
\bibitem[Bamba et al.(2003)]{2003ApJ...589..827B}
 Bamba, A., Yamazaki, R., Ueno, M., \& Koyama, K.\ 2003, \apj, 589, 827 
\bibitem[Bamba et al.(2005)]{2005ApJ...621..793B}
 Bamba, A., Yamazaki, R., Yoshida, T., et al.\ 2005, \apj, 621, 793 
\bibitem[Bamba et al.(2012)]{2012ApJ...756..149B}
 Bamba, A., P{\"u}hlhofer, G., Acero, F., et al.\ 2012, \apj, 756, 149 
\bibitem[Borkowski et al.(2001)]{2001ApJ...550..334B}
 Borkowski, K.~J., Rho, J., Reynolds, S.~P., \& Dyer, K.~K.\ 2001, ApJ, 550, 334
\bibitem[Broersen et al.(2014)]{2014MNRAS.441.3040B}
 Broersen, S., Chiotellis, A., Vink, J., \& Bamba, A.\ 2014, \mnras, 441, 3040 
\bibitem[Castro et al.(2013)]{2013ApJ...779...49C}
 Castro, D., Lopez, L.~A., Slane, P.~O., et al.\ 2013, \apj, 779, 49 
\bibitem[Foster et al.(2012)]{2012ApJ...756..128F}
 Foster, A.~R., Ji, L., Smith, R.~K., \& Brickhouse, N.~S.\ 2012, \apj, 756, 128 
\bibitem[Fraschetti et al.(2016)]{fraschetti2016}
Fraschetti et al.\ 2016, in prep.
\bibitem[Ghavamian et al.(2001)]{ghavamian2001}
Ghavamian, P., Raymond, J., Smith, R.~C., \& Hartigan, P.\ 2001, \apj, 547, 995 
\bibitem[Green(1988)]{1988Ap&SS.148....3G}
 Green, D.~A.\ 1988, Ap\&SS, 148, 3 
\bibitem[Inoue et al.(2013)]{2013ApJ...772L..20I}
 Inoue, T., Shimoda, J., Ohira, Y., \& Yamazaki, R.\ 2013, \apjl, 772, L20 
\bibitem[Ishisaki et al.(2007)]{2007PASJ...59S.113I}
 Ishisaki, Y., Maeda, Y., Fujimoto, R., et al.\ 2007, PASJ, 59, 113 
\bibitem[Helder et al.(2011)]{2011ApJ...737...85H}
 Helder, E.~A., Vink, J., \& Bassa, C.~G.\ 2011, \apj, 737, 85 
\bibitem[Helder et al.(2013)]{2013MNRAS.435..910H}
 Helder, E.~A., Vink, J., Bamba, A., et al.\ 2013, \mnras, 435, 910 
\bibitem[Hess(1912)]{Hess}
 Hess V.F. 1912, Phys. Zeits 13, 1084
\bibitem[Katsuda(2014)]{katsuda2014}
Katsuda, S.\ 2014, 40th COSPAR Scientific Assembly, 40, 1426
\bibitem[Katsuda et al.(2010)]{2010ApJ...723..383K}
 Katsuda, S., Petre, R., Mori, K., et al.\ 2010, \apj, 723, 383 
\bibitem[Katsuda et al.(2015)]{2015ApJ...814...29K}
 Katsuda, S., Acero, F., Tominaga, N., et al.\ 2015, \apj, 814, 29
\bibitem[Koyama et al.(1995)]{1995Natur.378..255K}
 Koyama, K., Petre, R., Gotthelf, E.~V., et al.\ 1995, \nat, 378, 255 
\bibitem[Koyama et al.(2007)]{2007PASJ...59S..23K}
 Koyama, K., Tsunemi, H., Dotani, T., et al.\ 2007, PASJ, 59, 23 
\bibitem[Long \& Blair(1990)]{long1990}
Long, K.~S., \& Blair, W.~P.\ 1990, \apjl, 358, L13 
\bibitem[Malkov \& Drury(2001)]{2001RPPh...64..429M}
Malkov, M.~A., \& Drury, L.~O.\ 2001, Reports on Progress in Physics, 64, 429 
\bibitem[Miceli et al.(2009)]{2009A&A...501..239M}
 Miceli, M., Bocchino, F., Iakubovskyi, D., et al.\ 2009, \aap, 501, 239
\bibitem[Mitsuda et al.(2007)]{2007PASJ...59S...1M}
 Mitsuda, K., Bautz, M., Inoue, H., et al.\ 2007, PASJ, 59, 1 
\bibitem[Nakajima et al.(2008)]{nakajima2008}
Nakajima, H., Yamaguchi, H., Matsumoto, H., et al.\ 2008, \pasj, 60, 1 
\bibitem[Naranan et al.(1977)]{1977ApJ...213L..53N}
 Naranan, S., Shulman, S., Yentis, D., Fritz, G., \& Friedman, H.\ 1977, ApJL, 213, L53 
\bibitem[Patnaude et al.(2009)]{2009ApJ...696.1956P}
 Patnaude, D.~J., Ellison, D.~C., \& Slane, P.\ 2009, \apj, 696, 1956 
\bibitem[Petruk et al.(2009)]{2009MNRAS.393.1034P}
 Petruk, O., Dubner, G., Castelletti, G., et al.\ 2009, \mnras, 393, 1034 
\bibitem[Rho et al.(2002)]{2002ApJ...581.1116R}
 Rho, J., Dyer, K.~K., Borkowski, K.~J., \& Reynolds, S.~P.\ 2002, \apj, 581, 1116 
\bibitem[Reynolds \& Keohane(1999)]{1999ApJ...525..368R}
 Reynolds, S.~P., \& Keohane, J.~W.\ 1999, \apj, 525, 368 
\bibitem[Reynoso et al.(2013)]{2013AJ....145..104R}
 Reynoso, E.~M., Hughes, J.~P., \& Moffett, D.~A.\ 2013, \aj, 145, 104 
\bibitem[Rosado et al.(1996)]{1996A&A...315..243R}
 Rosado, M., Ambrocio-Cruz, P., Le Coarer, E., \& Marcelin, M.\ 1996, A\&A, 315, 243 
\bibitem[Rothenflug et al.(2004)]{2004A&A...425..121R}
 Rothenflug, R., Ballet, J., Dubner, G., et al.\ 2004, \aap, 425, 121 
\bibitem[Sano et al.(2016)]{sano2016}
Sano, H., Nakamura, K., Furukawa, N., et al.\ 2016, arXiv:1606.07745
\bibitem[Serlemitsos et al.(2007)]{2007PASJ...59S...9S}
 Serlemitsos, P.~J., Soong, Y., Chan, K.-W., et al.\ 2007, PASJ, 59, 9 
\bibitem[Shimoda et al.(2015)]{shimoda2015}
Shimoda, J., Inoue, T., Ohira, Y., et al.\ 2015, \apj, 803, 98 
\bibitem[Slane et al.(2001)]{2001ApJ...548..814S}
 Slane, P., Hughes, J.~P., Edgar, R.~J., et al.\ 2001, ApJ, 548, 814 
\bibitem[Tawa et al.(2008)]{2008PASJ...60S..11T}
 Tawa, N., Hayashida, K., Nagai, M., et al.\ 2008, PASJ, 60, 11 
\bibitem[Takahashi et al.(2008)]{2008PASJ...60S.131T}
 Takahashi, T., Tanaka, T., Uchiyama, Y., et al.\ 2008, \pasj, 60, 131 
\bibitem[Uchiyama et al.(2009)]{2009PASJ...61S...9U}
 Uchiyama, H., Ozawa, M., Matsumoto, H., et al.\ 2009, PASJ, 61, 9 
\bibitem[Vink et al.(2006)]{2006ApJ...648L..33V}
 Vink, J., Bleeker, J., van der Heyden, K., et al.\ 2006, \apjl, 648, L33 
\bibitem[Yamaguchi et al.(2008)]{2008PASJ...60S.123Y}
 Yamaguchi, H., Koyama, K., Nakajima, H., et al.\ 2008, PASJ, 60, 123 
\bibitem[Yamaguchi et al.(2011)]{2011PASJ...63S.837Y}
 Yamaguchi, H., Koyama, K., \& Uchida, H.\ 2011, \pasj, 63, 837 
\bibitem[Yamaguchi et al.(2016)]{yamaguchi2016}
Yamaguchi, H., Katsuda, S., Castro, D., et al.\ 2016, \apjl, 820, L3 
\bibitem[Yamazaki et al.(2006)]{yamazaki2006}
Yamazaki, R., Kohri, K., Bamba, A., et al.\ 2006, \mnras, 371, 1975 
\bibitem[Yang et al.(2014)]{2014A&A...567A..23Y}
 Yang, R.-z., Zhang, X., Yuan, Q., \& Liu, S.\ 2014, \aap, 567, AA23 
\end{thebibliography}
\end{document}